\documentstyle{mn}

\newif\ifAMStwofonts

\ifoldfss
  \ifCUPmtlplainloaded \else
    \NewTextAlphabet{textbfit} {cmbxti10} {}
    \NewTextAlphabet{textbfss} {cmssbx10} {}
    \NewMathAlphabet{mathbfit} {cmbxti10} {} 
    \NewMathAlphabet{mathbfss} {cmssbx10} {} 
  \fi
  \ifAMStwofonts
    \ifCUPmtlplainloaded \else
      \NewSymbolFont{upmath} {eurm10}
      \NewSymbolFont{AMSa} {msam10}
      \NewMathSymbol{\upi}     {0}{upmath}{19}
      \NewMathSymbol{\umu}     {0}{upmath}{16}
      \NewMathSymbol{\upartial}{0}{upmath}{40}
      \NewMathSymbol{\leqslant}{3}{AMSa}{36}
      \NewMathSymbol{\geqslant}{3}{AMSa}{3E}

      \let\leq=\leqslant 
       
    \fi
  \fi
\fi 

\ifnfssone
  \newmathalphabet{\mathit}
  \addtoversion{normal}{\mathit}{cmr}{m}{it}
  \addtoversion{bold}{\mathit}{cmr}{bx}{it}
  \newmathalphabet{\mathbfit} 
  \addtoversion{normal}{\mathbfit}{cmr}{bx}{it}
  \addtoversion{bold}{\mathbfit}{cmr}{bx}{it}
  \newmathalphabet{\mathbfss} 
  \addtoversion{normal}{\mathbfss}{cmss}{bx}{n}
  \addtoversion{bold}{\mathbfss}{cmss}{bx}{n}
  \ifAMStwofonts
    \ifCUPmtlplainloaded \else
      \UseAMStwoboldmath
      \makeatletter
      \new@mathgroup\upmath@group
      \define@mathgroup\mv@normal\upmath@group{eur}{m}{n}
      \define@mathgroup\mv@bold\upmath@group{eur}{b}{n}
      \edef\UPM{\hexnumber\upmath@group}
      \new@mathgroup\amsa@group
      \define@mathgroup\mv@normal\amsa@group{msa}{m}{n}
      \define@mathgroup\mv@bold\amsa@group{msa}{m}{n}
      \edef\AMSa{\hexnumber\amsa@group}
      \makeatother
      \mathchardef\upi="0\UPM19
      \mathchardef\umu="0\UPM16
      \mathchardef\upartial="0\UPM40
      \mathchardef\leqslant="3\AMSa36
      \mathchardef\geqslant="3\AMSa3E

      \let\leq=\leqslant 

    \fi
  \fi
\fi 

\ifnfsstwo
  \DeclareMathAlphabet{\mathbfit}{OT1}{cmr}{bx}{it}
  \SetMathAlphabet\mathbfit{bold}{OT1}{cmr}{bx}{it}
  \DeclareMathAlphabet{\mathbfss}{OT1}{cmss}{bx}{n}
  \SetMathAlphabet\mathbfss{bold}{OT1}{cmss}{bx}{n}
  \ifAMStwofonts
    \ifCUPmtlplainloaded \else
      \DeclareSymbolFont{UPM}{U}{eur}{m}{n}
      \SetSymbolFont{UPM}{bold}{U}{eur}{b}{n}
      \DeclareSymbolFont{AMSa}{U}{msa}{m}{n}
      \DeclareMathSymbol{\upi}{0}{UPM}{"19}
      \DeclareMathSymbol{\umu}{0}{UPM}{"16}
      \DeclareMathSymbol{\upartial}{0}{UPM}{"40}
      \DeclareMathSymbol{\leqslant}{3}{AMSa}{"36}
      \DeclareMathSymbol{\geqslant}{3}{AMSa}{"3E}

      \let\leq=\leqslant 

    \fi
  \fi
\fi 

\ifCUPmtlplainloaded \else
  \ifAMStwofonts \else 
    \def\upi{\pi}
    \def\umu{\mu}
    \def\upartial{\partial}
  \fi
\fi

\title{The discovery of low-mass pre main sequence stars in Cepheus OB3b}
\author[M. Pozzo, T. Naylor, R.D. Jeffries, J.E. Drew]
       {M. Pozzo$^{1,3}$, T. Naylor$^{2,3}$\thanks{Guest investigator of the UK Astronomical
        Data Centre.}, R.D. Jeffries$^3$, J.E. Drew$^1$ \\
        $^1$ Imperial College London, Blackett Laboratory,
        Prince Consort Road, London, SW7 2BW, U.K. \\
        $^2$ School of Physics, University of Exeter, Stocker Road, Exeter, EX4 4QL, U.K.\\
        $^3$ Astrophysics Group, School of Chemistry and Physics, Keele University, Staffordshire, ST5 5BG, U.K.}

\date{}
       
\pagerange{\pageref{firstpage}--\pageref{lastpage}}
\pubyear{0000}

\begin{document}

\maketitle

\label{firstpage}

\begin{abstract}
We report the discovery of a low-mass pre-main-sequence (PMS) stellar
population in the younger subgroup of the Cepheus OB3 association, Cep
OB3b, using {\it UBVI} CCD photometry and follow-up spectroscopy.  The
optical survey covers about 1300 square arcmin on the sky and gives a
global photometric and astrometric catalogue for more than 7000
objects.  The location of a PMS is well defined in a V versus (V-I)
colour-magnitude diagram (CMD).
 
Multi-fibre spectroscopic results for optically-selected PMS candidates
confirm the T Tauri nature for 10 objects, with equal numbers of 
classical TTS (CTTS) and weak-line TTS (WTTS).
There are 6 other objects which we classify as possible PMS stars.  The
newly discovered TTS stars have masses in the range $\sim 0.9 - 3.0$
M$_{\odot}$ and ages from $<1$ to nearly 10 Myr, based on the Siess, Dufour
\& Forestini (2000) isochrones.  Their location close to the O and B
stars of the association (especially the O7n star) demonstrates that
low-mass star formation is indeed possible in such an apparently
hostile environment dominated by early-type stars and that the latter
must have been less effective in eroding the circumstellar discs of their
lower mass siblings compared to other OB associations (e.g., $\lambda$
Ori).  We attribute this to the nature of the local environment,
speculating that the bulk of molecular material, which shielded
low-mass stars from the ionising radiation of their early-type siblings,
has only recently been removed.

\end{abstract}

\begin{keywords}
open clusters and associations: individual: Cepheus OB3 - stars: pre-main-sequence 
- stars: formation - stars: low-mass, brown dwarfs - accretion, accretion discs 
- techniques: radial velocities - techniques: spectroscopic
\end{keywords}

\section{Introduction}

In the last few years it has become increasingly recognised that the
majority of low-mass stars in the solar neighbourhood are likely to
have formed in OB associations.  From extinct radionuclide abundances,
it also seems plausible that our Sun was born close to an OB
association (see Harper 1996, and references therein).  OB associations
are therefore key to understanding star formation processes and testing
the universality of the initial mass function (IMF). From recent
studies, the number of low-mass stars formed seems to follow the
Miller-Scalo IMF of the field stars (see for example Walter \& Boyd
1991; Walter et al. 1994; Preibisch \& Zinnecker 1999; Preibisch,
Guenther \& Zinnecker 2001; Dolan \& Mathieu 2001).

Although OB associations are by definition unbound, they are young
enough that dynamical effects such as mass segregation and preferential
evaporation of lower mass stars (de la Fuente Marcos 1995) will not
have occurred.  It is crucial to establish the influence that high-mass
neighbours have on the formation and evolution of their lower mass
siblings. The winds and ionising radiation of hot stars could influence
the mass function and circumstellar disc lifetimes of the lower mass
stars, with implications for angular momentum evolution and planet
formation. These ideas have gained currency with the discovery of
evaporating discs around PMS stars in the Orion nebula (McCaughrean \&
O'Dell 1996) and theoretical studies showing that discs could be
ionised and evaporated by the UV radiation fields of O stars
(Johnstone, Hollenbach \& Bally 1998).

The fact that the majority of low-mass stars are likely to form in
high-mass OB associations has not inspired a huge number of
observational projects to look for them, because of their faintness and
the large association dimensions.  Objective-prism H$\alpha$ studies
were the first tool used to identify young low-mass PMS stars. These
were mainly classical T~Tauri stars (CTTS) which show the strongest
H$\alpha$ emission (see reviews by Herbig 1962; Bastian et al. 1983;
Bertout 1989; Appenzeller \& Mundt 1989).  These studies were followed
by X-ray surveys, which have proved to be a convenient method to find
mainly the more weakly H$\alpha$ emitting, so-called weak-line T~Tauri
stars (WTTS; see discussion in Stelzer \& Neuh{\"a}user 2001): these
seem to be stronger soft 
X-ray emitters than CTTS, perhaps because of their faster rotation
(see Bouvier et al. 1993, 1995; Wichmann et al. 1998) or perhaps
because a large fraction of the X-ray flux is likely to be absorbed by the
optically thick circumstellar material around CTTS (see Gahm 1981;
Walter \& Kuhi 1981; Flaccomio, Micela \& Sciortino 2003a).  Besides
such a bias against CTTS, there is also a more general possible incompleteness
problem: Wolk (1996) found X-ray selected PMS stars sharing the same
position in the CMD of non X-ray selected PMS stars -- demonstrating the
presence of less magnetically active PMS stars. Recent surveys by
more sensitive X-ray satellites such as {\em Chandra} and {\em
XMM-Newton} can mitigate against both these biases, if such surveys
are deep enough. For instance a very large fraction of optical and IR
low-mass members of the Orion nebula cluster were detected as {\em Chandra}
X-ray sources (Feigelson et al. 2002; Flaccomio et al. 2003b).

With the advent of larger-area sensitive detectors, it has become
possible to undertake photometric studies of OB associations covering
large areas with CCD mosaics, yielding
identifications of the faintest low-mass members (e.g., Pozzo et
al. 2000).  When coupled with follow-up spectroscopy (e.g., Dolan \&
Mathieu 1999, 2001; Rebull et al. 2000; Preibisch et al. 2001) these
surveys can overcome any possible biases associated with X-ray selected samples
(Walter et al. 2000).  This is the method we used for the investigation
of our target region, the Cepheus OB3 association.

\begin{figure}
\vspace{8.7cm}
\includegraphics{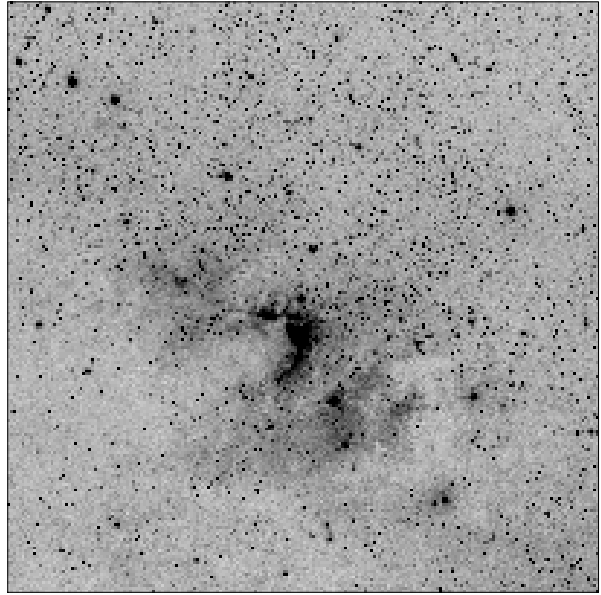}
\caption[The Palomar Sky Survey image of Cep OB3b]
{The Palomar Sky Survey image ($\sim 1.7^{\circ} \times 1.7^{\circ}$) of the region 
surrounding the Cep OB3b subgroup.}
\label{palomar}
\end{figure}

Cepheus OB3 is a very young association at about $800$ pc from the Sun 
(Moreno-Corral et al. 1993), with galactic coordinates 
{\it l}$^{II}=109.3^{\circ}$ - $111.7^{\circ}$ and {\it b}$^{II}=2.4^{\circ}$ 
- $3.7^{\circ}$ (Blaauw 1964). It covers a region on the sky from about
$22^h 46^m$ to $23^h 10^m$ in right ascension, and from about $+61^{\circ}$ 
to $+64^{\circ}$ in declination. 
It is composed of two main subgroups: the older, Cep OB3a, 
whose largest projected dimension is 17 pc; and the younger, Cep OB3b (shown in 
the Palomar Sky Survey image in Fig. \ref{palomar}), more compact and closer to the 
molecular cloud, whose largest projected dimension is 10 pc (Blaauw 1964). 

Blaauw, Hiltner \& Johnson (1959, and {\it erratum} in 1960; hereafter BHJ)
identified 40 association members in total, comprising three late O-type and 
33 B-type stars; some PMS members were proposed by Garmany (1973) 
and Sargent (1979); 8 new association members were recently proposed by 
Jordi, Trullols \& Galad\`\i-Enr\`{\i}quez (1996; hereafter J96).

The ages of Cep OB3a and Cep OB3b were estimated by isochrone fitting
to the high-mass stars to be about $10\pm2$ Myr and $7\pm2$ Myr
respectively (see Blaauw 1964, 1991; de Zeeuw \& Brand 1985). Ages
derived from such isochrone fitting must be treated with caution in
young OB associations: a better way is to look for low-mass stars which
are still contracting towards the main-sequence.  Note, however, that
masses and ages remain model dependent.  As a consequence, the shape of
both the evolutionary tracks and of the isochrones change from one
model to another, leading to discrepancies in the age estimates.

Much lower values of $\simeq 0.5$ Myr can be found in the literature
for ages derived using the expansion age (kinematic methods) of the two
Cep OB3 subgroups separately, or the association as a whole (De Vegt
1966; Garmany 1973; Assousa, Herbst \& Turner 1977; Sargent 1979;
Trullols et al. 1997). Compared with the much larger nuclear ages from
isochrone fitting there is an apparent paradox. However, the kinematic
methods have been shown most commonly to underestimate the true age
(i.e., the nuclear age; Brown, Dekker \& de Zeeuw 1997).  Brown et
al. (1997) advise that a kinematic age of $\sim$ 0.5 Myr for Cep OB3
``should be treated with caution''.

What was the trigger which initiated star formation in Cep OB3 (i.e,
the older subgroup Cep OB3a)? And what was subsequently responsible for
the creation of the younger subgroup Cep OB3b? Are there signs of
further star formation processes indicative of a third generation of
newly born stars in the association?  The members of the younger
subgroup lie between the older subgroup and the molecular cloud.  In
the darkest part of the cloud, $A_V= 46$ (Felli et al. 1978), and so
any optical detection of embedded stars is impossible.  The fact that
there is a temporal separation between the older and the younger
subgroup of about 3 Myr as well as a spatial separation of about 13~pc
strengthened the hypothesis of the sequential model (Elmegreen \& Lada
1977) for star formation in the Cep OB3 association (see Sargent 1979).
According to this model, the ionising radiation of early-type stars is
responsible for the propagation of a shock wave front through the
cloud, making the material gravitationally unstable and causing its
condensation into new massive stars. In other words, once a subgroup is
formed, another one grows thanks to shocks driven by the previous one.

A different scenario was instead proposed by Assousa et al. (1977),
who confirmed the existence of an H~{\small I} expanding shell 
around Cep OB3, first discovered by Simonson \& van Someren Greve (1976)  
from 21-cm radio line observations. The H~{\small I} shell was found to be centred on
the older subgroup (NE of the younger subgroup), with a radius of
about 53 pc and expanding with a velocity of 35 km s$^{-1}$.  The shell
was identified with a supernova (SN) remnant (type II) of age 0.43\,Myr.
The pulsar PSR 2223+65, 1.14\,Myr old,
was suggested as a possible stellar remnant of the SN event. In light
of these results, Assousa et al. (1977) proposed a SN-induced star
formation process for the younger subgroup, Cep OB3b, which was then
estimated to be of age 0.3\,Myr by kinematic methods. We
know now that Cep OB3b is instead likely to be a few Myr old, implying
that the proposed pulsar is too young to be the remnant of the SN
triggering star formation in this subgroup. However, another pulsar
quoted by Assousa et al. (1977), PSR 2324+60, some 10 Myr old, may be a
better candidate for having indeed triggered star formation not only in
the younger subgroup but, possibly, in the older subgroup too.

Finally, the geography of the region must be considered in detail.  The
molecular cloud is in physical contact with the S155 H~{\small II} region 
(see Panagia \& Thum 1981), about 35 pc in size (Heyer, Carpenter \& Ladd 1996), 
which was created by the ionising radiation of the OB stars. A bright arc of 
nebulosity defines the interface between the S155 H~{\small II} region and the
molecular cloud (see Fig. \ref{palomar}).  The presence of an H~{\small
II} region and of a reflection nebula, i.e. the signs of the
interaction between the molecular cloud and the OB stars previously
formed in the association, confirm that recent star formation has taken
place in the region.  This is supported by the radio and near-infrared
survey results of Testi et al. (1995), suggesting the presence of a
cluster of young embedded stars which could represent the third
generation of early-type stars in Cep OB3.  Recently, Naylor \&
Fabian (1999) have discovered more than 50 X-ray sources in Cep OB3b
thanks to {\it ROSAT} observations: their X-ray properties strongly
suggest their likely T Tauri nature, corroborating the hypothesis that
the presence of high-mass stars does not halt low-mass star formation, as
previously widely thought.  Furthermore, the discovery of another
compact H~{\small II} region, south of Cep OB3b (Harten, Thum \& Felli
1981), picks out another possible area of active star formation. Given
these results, we decided to investigate this long-studied
association, concentrating our attention on the younger subgroup, Cep
OB3b.

In this paper we present and discuss an {\it UBVI}\, CCD photometric
study of the Cep OB3b subgroup. For the first time, thanks to the wide
surveyed area (of about 1300 arcmin$^2$) and deep photometry, we have
been able to unambiguously detect low-mass pre-main sequence members of
the Cep OB3 association.  An optically-selected sample of PMS
candidates was spectroscopically followed-up at optical wavelengths, to
test for kinematic membership of the association and confirm youth
through radial velocity and Li\,{\sc i} 6708 \AA\, equivalent width
measurements.

The paper is organised as follows. In Section 2, photometric
observations and the data reduction are presented; colour-magnitude
diagrams for each of the observed fields are described in Section
3. Section 4 deals with the investigation of the stars in the
colour-magnitude diagrams by means of the $Q$-method and by
cross-correlation of our optical photometric catalogue with the {\it
ROSAT} X-ray selected young star candidates presented by NF99.  In
Section 5 we discuss the results obtained from a spectroscopic
follow-up of our photometry: using radial velocity measurements,
spectral classification and Li\,{\sc i} 6708 \AA\, equivalent widths we
identify PMS objects in Cep OB3b.  Possible age gradients of PMS
objects suggested by isochrone fitting are studied in Section 6.
Discussion is presented in Section 7 and conclusions follow in Section
8.

\section{Optical photometric observations and data reduction}\label{datared}
 
\begin{figure}
\vspace{7cm}
\includegraphics{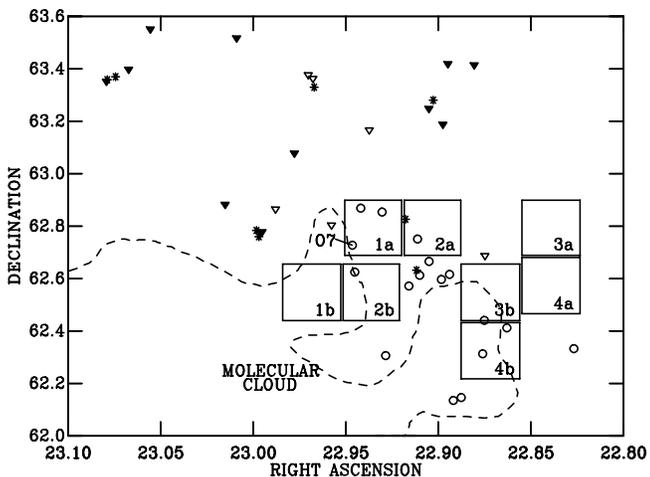} 
\caption[]
{Our surveyed region in the Cep OB3 association. The superimposed boxes are our survey 
CCD fields for CCD2 (fields a) and CCD4 (fields b). 
BHJ members of the Cep OB3a subgroup are shown as filled triangles, 
those of Cep OB3b as open circles; possible PMS stars from Garmany (1973) and Sargent (1977) 
are shown as open triangles; new possible members (J96) as asterisks; 
the O7n star (BHJ 41) is marked; the dashes represent the lowest $^{12}$CO contour of 
peak antenna temperatures for the molecular cloud (taken from Sargent 1977).}
\label{figfields3}
\end{figure}

The {\it UBVI} CCD images were taken on the nights of August 10/11 and 11/12
1997, with the 2.5-m Isaac Newton Telescope (INT), La Palma, Canary
Islands, Spain, equipped with the Wide Field Camera (WFC) and an array of
4 Loral CCDs. Data were taken using filters $U$ (RGO), $B$ (Harris), $V$ (Harris)
and $I$ (RGO INT-WFC). 
Even though the WFC was composed of four Loral CCDs, only data from CCD2 and CCD4 
were reduced: CCD1 was dead, while CCD3 had a sensitivity problem and a very
non-uniform response. The images cover approximately $12 \times 12$ arcmin$^2$ each (one
pixel is equivalent to about 0.37 arcsec on the sky). Field centres and exposure times
are listed in Table 1. Fig. \ref{figfields3} shows the surveyed region.
The average seeing was $1.3$ arcsec during the first night of observation, and $1.1$ arcsec 
in the second night. Both nights were photometric.
   
After the images had been debiassed and flat fielded, we extracted the flux
for each star in the field using the optimal extraction techniques outlined
in Naylor (1998) and Naylor et al. (2002).
Our final catalogue gives V,  (U-B), (B-V) and (V-I) for each star along
with an associated error and a flag.
Non-zero flags are used to report problems with particular stars which make
the data unreliable, details of which are given in Naylor et al. (2002).
The catalogue also contains astrometric positions for each star, using
a slightly modified version of the procedure described in Naylor et al. (2002).
All of fields 1a, 2a and 2b, and 80 percent of 1b are covered by the second 
incremental release of the 2 Micron All-Sky Survey (2MASS) catalogue.
So we began fitting our X-Y positions to the 2MASS positions, using 
a nine co-efficient model, which included the tangent point and a 
pincushion co-efficient as free parameters.
The mean pincushion co-efficient was consistent with the nominal value of 
220 rad$^{-2}$ for the INT prime focus, so this was adopted for all the fields.
We then re-fitted all the CCDs, this time using the (less accurate) USNO A-2.0 
catalogue positions (Monet 1998) for the fields where 2MASS was not available, 
to derive the mean tangent point for each pointing by averaging the value for the
two CCDs. This allowed a final six co-efficient model to be fitted, using (for 
consistency) USNO A-2.0 for all fields.
The RMS residual about the fits was about 0.35 arcsec, though more in fields
4b and 1a (0.43 and 0.53 arcsec respectively) where there were a shortage of
USNOA-2.0 stars.

\begin{table}
 \centering
 \begin{minipage}{82mm}
  \caption{CCD fields surveyed in the Cep OB3 region. Fields a correspond to CCD2, fields b to CCD4 (see Figure \ref{figfields3}).} 
  \begin{tabular}{@{}ccc@{}}
CCD & centre (RA, dec) & Exposure times (sec)  \\
fields &  (J2000.0)    &   in U,B,V,I (long; short) \\[10pt] 
1a   & 22 56 06.78 \, +62 47 31.67  & 100, 40, 20, 10; 10, 4, 1, 1 \\
1b   & 22 58 05.34 \, +62 32 41.21  & 100, 40, 20, 10; 10, 4, 1, 1 \\
2a   & 22 54 11.77 \, +62 47 42.84  & 100, 40, 20, 10; 10, 4, 2, 1 \\
2b   & 22 56 10.43 \, +62 32 52.44  & 100, 40, 20, 10; 10, 4, 2, 1 \\
3a   & 22 50 21.14 \, +62 47 44.88  & 100, 40, 20, 10; 10, 4, 2, 1 \\
3b   & 22 52 19.63 \, +62 32 53.92  & 100, 40, 20, 10; 10, 4, 2, 1 \\
4a   & 22 50 21.78 \, +62 34 22.50  & 100, 40, 20, 10; 10, 4, 1, 1 \\
4b   & 22 52 18.97 \, +62 19 32.20  & 100, 40, 20, 10; 10, 4, 1, 1 \\
\end{tabular}
\end{minipage}
\end{table}

As explained in Naylor et al. (2002), 
to derive the magnitudes in the catalogue, we need a set of photometric
co-efficients derived using a relatively large aperture.
To achieve this we observed one field in the older subgroup (termed the 
local field) at four different times during the run.
We then measured bright stars in this field with varying size apertures,
and found that the smallest scatter between measurements was achieved using a
radius of 15 pixels, or $5.6$ arcsec.
We then used this aperture to measure a total of 48 standard stars 
from three Landolt (1992) standard fields which were observed over both 
nights for both CCDs.
To derive the photometric co-efficients from these measurements we assumed
(i) the colour terms were the same on both nights, but were different for each
CCD; (ii) the extinction coefficient was the same for both CCDs, 
but different on each night; and (iii) the zero points were different for 
the two CCDs and from night to night.
This required a fit to $8$ parameters, (i.e., 2 colour terms, 4 zero points 
and 2 extinction coefficients), for each colour transformation equation. 
The fit is based on a weighted least-squares method. 
A systematic error was added in quadrature to the total error in order to find 
a reduced $\chi^2$ equal to $1$. 
The additional systematic errors were about $1$ per cent for the
(V-I), (B-V) and V calibrations, but about $2.7$ per cent for (U-B), as we 
expected, since the U-band is always most problematic for CCD observations.
Multiple observations of the local field allowed us to test our final
field-to-field internal precision, which was about 0.02 mag.

We removed from the catalogue those stars for which no (V-I) colour or visual magnitude 
was determined to better than 0.1 mag and/or with corresponding non-zero flags. 
The resulting catalogue contained 7315 objects, with visual magnitudes
in the range $12.32\leq V \leq 21.04$ mag and colours in the range
$0.02\leq V-I \leq 6.58$ (on the Cousins system). 
This will be the only catalogue used in the subsequent analysis, unless otherwise specified.
The photometry has uncertainties smaller than 0.1 mag in all the {\it UBVI}
colours for 1241 objects, with visual magnitudes in the range
$V=12.32-19.88$ mag. Limiting magnitudes for each filter
(corresponding to an error of 0.1 mag) are about
$U=22.4$, $B=21.8$, $V=21.0$ and $I=20.0$.

For completeness, we have correlated our optical photometric catalogue
with the much smaller one of Jordi et al. (1995) in Appendix A.  We
found evidence for only small systematic discrepancies.

\section{Colour-magnitude diagrams}\label{cmdsec}

\begin{figure*}
\vspace{22.5cm}
\includegraphics{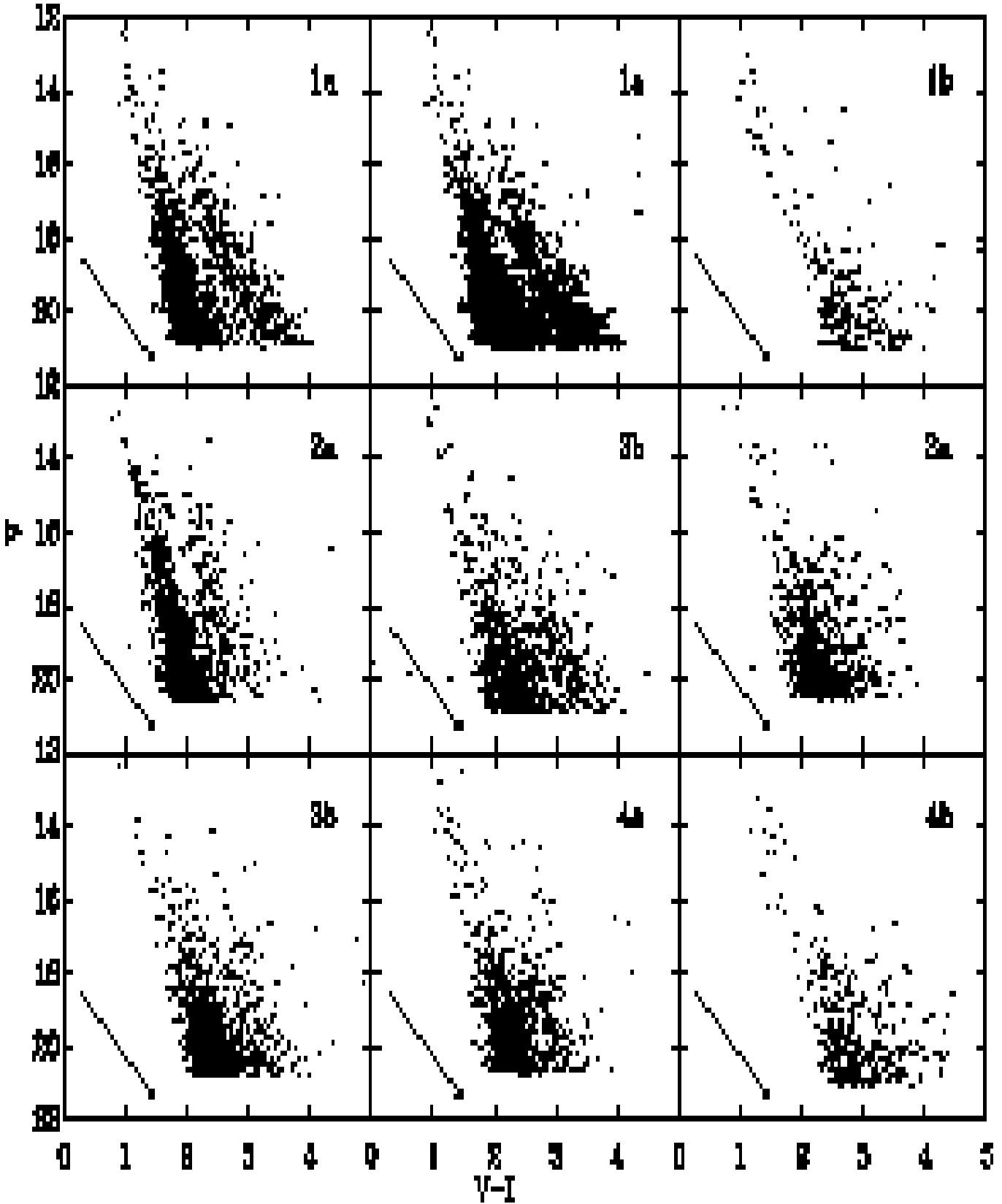}
\caption[]
{Colour-magnitude diagrams in V versus (V-I) for the surveyed fields as labelled.
The average reddening vector for Cep OB3b is shown, with $<A_V>=2.81$ and 
$E(V-I)=1.18$ (see Section \ref{cmdsec}). Also plotted (top-centre) there are
error bars for the field 1a data points (error bars for V = 12, 17, 20 are shown
as reference).}
\label{plot_9panels}
\end{figure*}

In Fig. \ref{plot_9panels}, the colour-magnitude diagrams (CMDs) in V versus
(V-I) for all the CCD fields are shown. Also plotted there is the mean
reddening vector we have adopted for the Cep OB3b subgroup, with
$<A_V>=2.81 \pm 0.10$ and $E(V-I)=1.18 \pm 0.03$, derived from
the relation $E(V-I)=1.3*E(B-V)$, taken from the reddening
relationship of Winkler (1997), with $E(B-V)=0.91 \pm 0.02$.
These values, and the mean distance modulus ($dm=9.65 \pm 0.20$ mag,
corresponding to a distance of about $850 \pm 80$ pc) were determined
from data given by Moreno-Corral et al. (1993). They derived the
extinction law for each of 14 OB stars individually, obtaining a mean R
($=A_V/E(B-V)$) value of $3.1 \pm 0.1$. The $A_V$ values for these stars range from
2.37 to 4.27 mag, while R ranges from 2.44 to 3.61.  We omitted BHJ
11 from the mean reddening determination because its membership is
doubtful, due to its position on the H-R diagram and its extinction,
$A_V=4.3$, which both differ from those of the confirmed Cep OB3
members (see Sargent 1979).  However, we point out that its inclusion
in the calculation would have had little effect, the slope of the
reddening vector staying practically the same in the CMDs.

Even though $R < 3.6$ for the early-type stars, it is possible that these are not
entirely representative of the whole Cep OB3b region: given that the
majority of the members are still close to, or embedded, in the
molecular cloud, it would not be surprising if R were larger for some
fraction of the subgroup members. Indeed, it has been shown that R can be
as high as 5 in the neighbourhood of star forming regions (Mathis
1990).  In other words, we do not necessarily 
expect R to have an homogeneous value
across the younger subgroup. A larger R value would
steepen the inclination of the reddening vector in the CMDs. For 
R$\simeq 5$ and our preferred $A_V=2.81$, $E(V-I)$ would
{\it decrease} by $\sim 0.2$.  In our later comparison of isochrones,
reddened according to the adopted mean extinction (see Section
\ref{isofit}), stars with an R value higher than the adopted 3.1 will
in fact be younger than they appear.

\begin{figure}
\vspace{7cm}
\includegraphics{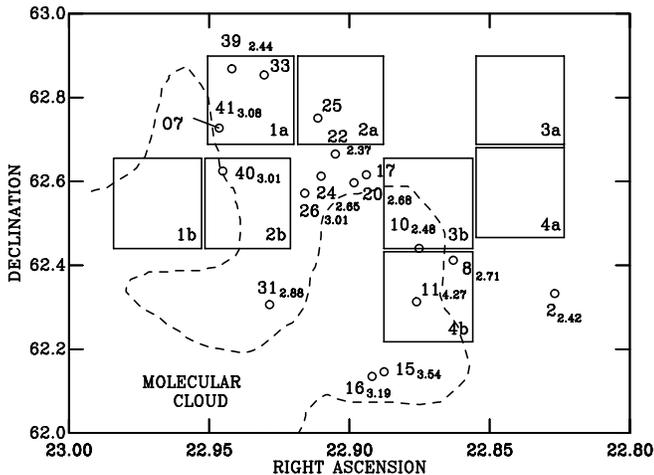} 
\caption[]
{The Cep OB3b subgroup, with known members shown as circles 
and their corresponding BHJ number (other symbols as in Fig. \ref{figfields3}).  
Those objects used to determine the mean reddening value have also a subscript 
indicating their $A_V$ value given by Moreno-Corral et al. (1993).
The $A_V$ value for BHJ 11 is also given, although not used in our calculation.}
\label{figfields3zoom}
\end{figure}

Fig. \ref{figfields3zoom} shows the location of the Cep OB3b members,
with their associated BHJ numbers: the stars which were used by us for
the determination of the mean reddening value have their $A_V$ value as
subscript (we give this value for BHJ 11 too, although not used in the
calculation).

\section{What are the stars in the colour magnitude diagrams?}\label{secstars}

\subsection{$Q$-method: the background population}

Let us analyse the CMDs in V versus (V-I) shown in Fig.
\ref{plot_9panels}, obtained for each of the 8 CCD fields,
in order to begin to identify the stars within them. 

The first thing which catches the eye is that in the CMDs of fields 1a,
2a and 2b (see Fig. \ref{plot_9panels}) there are two clear
sequences. The bluer of the two is the more densely populated in all
three fields.  By contrast, in the CMD of field 1b
(see Fig. \ref{plot_9panels}, top-right) most of this bluer sequence is
missing. Field 1b is directly towards the molecular cloud where we
would expect to see mainly foreground stars and slightly reddened stars
associated with it. We therefore conclude that the blue sequences
observed off-cloud in fields 1a, 2a and 2b are due to a background
population.  The separation of the redder sequences from these blue
sequences is not the product of random errors in the photometry. As an
example illustrating this, the plot given in Fig. \ref{plot_9panels},
top-centre, shows the data for field 1a with associated error-bars.  We
cannot identify any likely systematic error in the photometry that
could have led to an artificial division of the stars into two
sequences.
 
\begin{figure}
\vspace{7cm}
\includegraphics{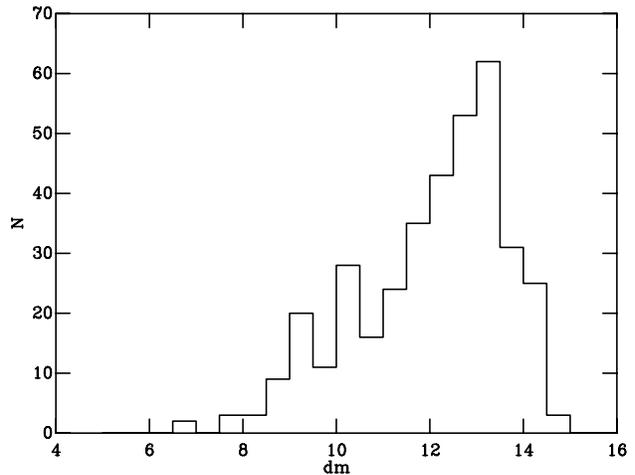}
\caption[Number of stars per $dm$]
{Number of stars per $dm$: the peak for Cep OB3b should be at about 9.6 mag. The considerable
peaks at larger distance moduli show the presence of background populations.}
\label{perseus}
\end{figure}

To gain an impression of the distance to the background objects we
consider field 1a, where the separation of the two sequences is
particularly clear.  A first guess can be made if we plot the number of
stars as a function of the distance modulus $dm$ ($=V-M_V-A_V= 5logd
-5$).  To do this we exploit the $Q$-method (Johnson \& Morgan, 1953)
which can uniquely find possible main-sequence members earlier than A5
(with luminosity classes III and V).  This is because for these stars
there is a unique {\it reddening independent} parameter
$Q=(U-B)-0.72*(B-V)$, which is less than 0.47 for spectral types
earlier than A5.  $Q$ permits us to determine the intrinsic $(B-V)$,
according to the relation $(B-V)_0= -0.009 + 0.337*Q$. From
intrinsic and measured $(B-V)$ colours we can then determine the colour
excess $E(B-V)$ and therefore $A_V$ according to the relation
$A_V=R*E(B-V)$ (assuming $R=3.1$).  The absolute visual magnitude
$M_V$, and hence $dm=V-M_V-A_V$, can then be found using a standard
$M_V$-$(B-V)_0$ relation.  We used the absolute-magnitude versus
intrinsic $(B-V)$ calibration for the ZAMS stars given by Walker
(1985).

We plot in Fig. \ref{perseus} the distance modulus for all the stars in field 1a with 
$Q$ values $< 0.47$. Consistent with the $Q$ limit, only intrinsic $(B-V)$ values 
in the range $-0.32$ to $0.15$ are retained (see also Fitzgerald 1970). 
The great majority of the stars included in the histogram lie in the
top part of the blue sequence seen in the V vs (V-I) diagram.
Stars belonging to Cep OB3b should have $dm \sim 9.65$ mag. As we can see from the plot, 
the distribution only just begins at this distance modulus and stretches to much larger values,
with the mode at $13.0 < dm < 13.5$ mag, implying a distance range from
1 to 10 kpc.
It seems reasonable to interpret these objects belonging to the blue sequence as mainly
B stars (since A stars are more common but fainter), concentrated in spiral arms and
less than 100 Myr old. There seem to be only a few objects associated
with the Cep OB3b subgroup at $\sim 800$ pc.

\begin{table*}
 \centering
 \begin{minipage}{150mm}
  \caption{Optical counterparts to the {\it ROSAT} X-ray sources of NF99. The listed coordinates 
are from our optical photometric catalogue.} 
  \begin{tabular}{@{}clrcccccl@{}}
 X-ray id. \footnote{(P) is for PSPC, (H) for HRI.} & V (NF99) &  n.c. & 
sep. (arcsec) & Field  & RA (J2000) & Dec (J2000) & V & (V-I)  \\[10pt]
 14 (P) &            &  6 &  7.1 & 1a & 22 55 30.103 & +62 47 39.56 & 16.94 & 2.18  \\
 20 (P) &            &  1 &  7.5 & 1a & 22 56 01.546 & +62 47 47.12 & 17.94 & 2.74  \\
 24 (H) &            &  7 &  3.2 & 1a & 22 56 18.345 & +62 45 16.84 & 18.14 & 2.80  \\
 26 (P) &            &  7 &  3.8 & 1a & 22 56 20.132 & +62 52 23.16 & 16.79 & 2.44  \\
 27 (H) & 15.3       & 10 &  4.8 & 1a & 22 56 26.417 & +62 41 29.23 & 15.38 & 1.91  \\
 29 (P) & 15.5       &  6 &  7.4 & 1a & 22 56 29.678 & +62 45 28.68 & 15.66 & 1.96  \\ 
 34 (H) & 14.2       &  5 &  1.6 & 2b & 22 56 38.721 & +62 37 14.43 & 14.55 & 2.26  \\
 35 (H) & 14.9       &  7 &  3.7 & 1a & 22 56 39.087 & +62 45 09.36 & 14.91 & 1.89  \\
 44 (P) & 13.5 (GSC) &  3 &  6.0 & 1a & 22 56 56.417 & +62 52 42.45 & 14.36 & 1.48\footnote{This object is very close (8 pixels, corresponding to 2.96 arcsec on the sky) to another star not numbered in the catalogue: possible binary?}  \\
 48 (P) &            &  2 & 12.9 & 2b & 22 57 05.220 & +62 38 38.60 & 15.15 & 1.51  \\
\end{tabular}
\end{minipage}
\end{table*}

\begin{table*}
 \centering
 \begin{minipage}{130mm}
  \caption{Probability of finding a correlation (the brightest) with a
 shifted PSPC (P) and HRI (H) X-ray source and the expected number of
 spurious correlations as a function of $V$.}
  \begin{tabular}{@{}ccccccc@{}}
 range  &  Prob(P)  & Prob(H) & N(P) \footnote{The number of correlations with the original X-ray sources in the V range shown.} & N(H) & N$_{r}$(P) \footnote{The number of random correlations expected in the V range shown.}& N$_{r}$(H) \\[10pt]
 V $<$ 15.15 & 0.02 & 0.02 & 1 & 2 & 0 & 0 \\ 
 V $<$ 16.15 & 0.05 & 0.02 & 3 & 3 & 0 & 0 \\ 
 V $<$ 17.15 & 0.06 & 0.04 & 5 & 3 & 0 & 0 \\ 
 V $<$ 18.15 & 0.10 & 0.09 & 6 & 4 & 1 & 0 \\ 
\end{tabular}
\end{minipage}
\end{table*}

It is known that Cep OB3 is in the Cygnus-Orion arm, between the Sagittarius-Carina arm 
and the Perseus arm. 
According to Verschuur (1973), the latter extends in distance from about 2.5 to 4.0 kpc 
(corresponding to a reddening-free distance modulus of between 12 and 13 mag).
There is also evidence for the existence of a second arm beyond the Perseus arm extending 
to higher galactic latitudes, passing through the line of sight of Cep OB3b 
(Kimeswenger \& Weinberger 1989; Ungerechts, Umbanhowar \& Thaddeus 2000). 
In addition, Wouterloot et al. (1990) combined CO measurements with IRAS point source data 
to show that star formation is indeed occurring along the Cep OB3b sight-line out to 
$\sim$10 kpc (see their figure 4).  
Fig. \ref{perseus} confirms that there is a considerable assemblage of stars 
beyond Cep OB3b.  

This same background population is apparent as the blue sequence in
each of fields 2a and 2b.  We have already noted that this sequence is
essentially absent from field 1b, positioned directly on the molecular
cloud.  In fields 3a, 3b and 4a (see Fig. \ref{plot_9panels}), the
background population of stars is still apparent but there is not the
neat separation between it and the redder population.  This is
presumably because the background objects are shifted towards larger
(V-I) colours due to additional interstellar material in these
directions (see Appendix B).  Field 4b (see Fig. \ref{plot_9panels},
bottom-right) is transitional in character between these fields and the
on-cloud 1b field.

\newpage

\subsection{Cross-correlation with X-ray sources: hints of a pre-main sequence population}

We cross-correlated our optical catalogue with the (0.1-2.4 KeV) {\it
ROSAT} X-ray sources recently detected by Naylor \& Fabian (1999;
hereafter NF99).  The suggestion made by NF99, that these X-ray sources
are T Tauri stars (low-mass PMS stars are strong X-ray emitters; see
for example Feigelson \& DeCampli 1981; Strom et al. 1990; Wichmann et
al. 1997; Feigelson \& Montmerle 1999), is strongly supported by both
the ratio of X-ray to optical luminosity for the faintest objects of
about $10^{-1.5}$ (which is typical of very active stars; see Stocke et
al. 1983), and by their X-ray variability, with values typical of other
PMS associations (see Montmerle et al. 1983).  30 of the NF99 sources
fall within our CCD fields.

We determined the separations between the X-ray sources and possible
optical counterparts, and accepted correlations within 15 arcsec of
Position Sensitive Proportional Counter (PSPC) positions, and 7 arcsec
of the more accurate High Resolution Imager (HRI) positions. There were
8 X-ray sources with no optical counterparts within these errors
circles. When more than one optical counterpart was found, the
brightest was then chosen because X-ray luminosity is approximately
proportional to the optical luminosity (see Feigelson et al. 1993).  At
this stage, 12 more of the X-ray sources had to be dropped from the
sample because their optical counterparts suffered photometric problems
or errors in colours larger than the 0.1 mag acceptance limit.  This
left a total of 10 counterparts to the NF99 X-ray sources with good
photometry, as listed in Table 2.  X-ray source number and visual
magnitude from NF99 are given respectively in columns 1 and 2.  The
number of correlations with optical counterparts within our adopted
error circles is given in column 3.  Column 4 refers to
the separation to the brightest optical counterpart belonging to the
field specified in column 5 (as given in Fig. \ref{figfields3}).
Columns 6 - 9 list the positions, visual magnitudes and (V-I) colours
of the brightest optical counterparts. 

We also derived the number of random correlations to be expected among
those found.  For this purpose, we repeated the cross-correlation by
shifting the X-ray positions of NF99 by 20 arcsec north, south, west
and east, and retaining only acceptable correlations according the
criteria used before.  The probability of having a correlation with one
of the shifted X-ray sources in a specified V range is reported in
Table 3, for the PSPC and HRI respectively (columns 1-3).  Also
tabulated are the number of good correlations found with the original
X-ray sources (columns 4-5) and the number of random correlations
(columns 6-7) in the corresponding V range for PSPC and HRI
respectively.  We therefore expect at most one out of the 10 optical
counterparts to be a random correlation (more precisely, one out of the
6 which are correlated with a PSPC X-ray source).

We stress that although there are several (up to 10) optical
correlations within our adopted X-ray error circles, the vast majority
of these are very faint, not candidate PMS members of Cep OB3b and not
credible candidate active stars (their X-ray to optical flux ratio
would be much too large). Furthermore, the "spurious" correlation probabilities we show
are for the entire CMD. Spurious correlations are
most likely to be found at colours where the stellar density is highest
for each V range and so not near the putative Cep OB3b PMS we describe
below. The density of candidate PMS stars is not large enough to make
it likely that more than one such star contributes to each X-ray
source. Lastly, the spurious correlation probabilities were calculated
using a 7 and 15 arcsec X-ray error circle for the HRI and PSPC sources
respectively. All but one of the correlations (one of the
optically brightest -- see below) are within the central 50 per cent of their
respective X-ray error circles, reducing the likelihood of them being
random correlations by a further factor of two over the figures quoted
in Table~3. Whilst we are thus quite confident that the optical
counterparts we have selected are indeed X-ray sources, we anticipate
that the scheduled {\it Chandra} AO-4 observation of Cep OB3b (P.I. G. Garmire)
will add a great deal, both in terms of its greater sensitivity --
enabling the detection of many fainter, less magnetically active PMS
stars, and with its narrower point spread function -- allowing
unambiguous optical counterpart identification.

\begin{figure}
\vspace{7cm}
\includegraphics{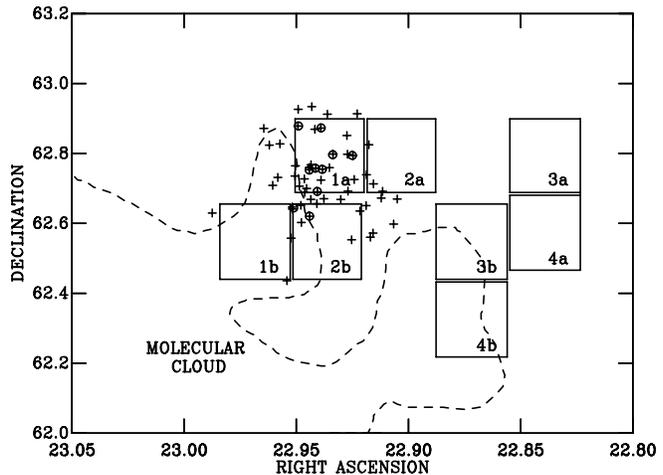}
\caption[The location of the 10 optical counterparts to the X-ray sources of NF99]
{The location of the 10 optical counterparts (circled) to the X-ray sources of NF99 
listed in Table 2. 
X-ray sources of NF99 for which we do not have an optical counterpart are simply shown
as crosses.}
\label{figfieldsXraycorr}
\end{figure}

\begin{figure}
\vspace{7cm}
\includegraphics{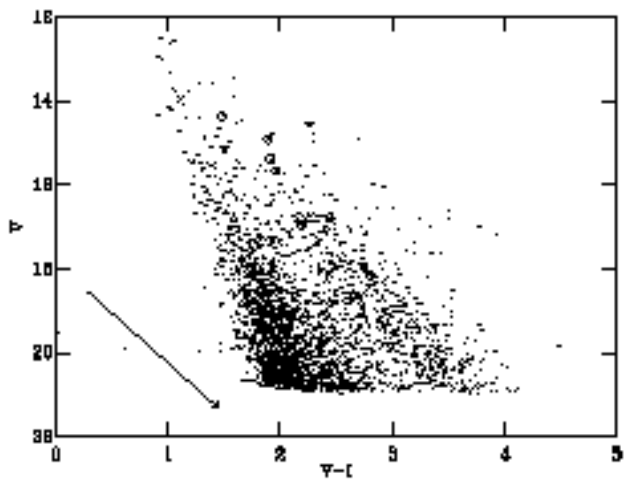}
\caption[The CMD for fields 1a and 2b, with the 10 optical counterparts to 
the X-ray sources of NF99]
{The CMD for fields 1a and 2b plotted together with the 10 
optical counterparts to the X-ray sources of NF99: those falling in the field 1a are 
shown as circles, whereas the two on field 2b are shown as open triangles. 
Also plotted there is the average reddening vector for Cep OB3b, with $<A_V>=2.81$ mag
and $E(V-I)=1.18$ mag (see Section \ref{cmdsec}).}
\label{plot1a2bXray}
\end{figure}

Fig. \ref{figfieldsXraycorr} shows the location of the 10 optical
counterparts to the X-ray sources of NF99 with respect to the CCD
fields.  All but two of them belong to field 1a, and lie in the redder
sequence of the CMD shown in Fig. \ref{plot1a2bXray}.  This is
consistent with the idea, but it does not yet prove, that this sequence
is a PMS, including young, active stars. There is still the possibility
of contaminating foreground and background X-ray active stars.
Note that the optical counterpart to one of the X-ray sources (no. 48)
in field 2b is the one closest to the ``background'' sequence in
Fig. \ref{plot1a2bXray}, at V=15.15 and (V-I)=1.51 and the most widely
separated counterpart from its X-ray source.  According to
the results of our spectroscopic follow-up (see Section
\ref{specclass}), it is an F-type star that cannot belong to Cep OB3b on
the grounds of its radial velocity.  This illustrates the essential
role for spectroscopy, which we now turn to, in finally establishing
the status of those stars belonging to the redder sequence in the CMDs.

\section{Spectroscopic follow-up}\label{specclass}

\subsection{Selection of PMS candidates}
 
Spectroscopic observations of the PMS objects found from our optical photometric survey 
are necessary to check their nature, i.e. to test their kinematic membership of the 
association and confirm that they are truly young objects. 
The first objective is achieved with radial velocity measurements and the the second can 
be achieved for cool stars by using the equivalent width of the Li\,{\sc i} 6708\AA\ line
as a youth discriminator (see Mart\'{\i}n 1997).

During the PMS phase, fully convective low-mass stars contract and their 
cores heat up to the point at which Li can be burned in p,$\alpha$ 
reactions. Convective mixing ensures that photospheric Li is then also 
depleted on very short timescales. In more massive stars, photospheric Li 
depletion occurs when and if the base of the convection zone achieves the 
threshold temperature for Li burning. Thus a Li abundance close to that of 
the pristine interstellar medium is a signature of youth in stars 
of low and intermediate masses. Unfortunately, both the time 
dependence of Li depletion and observational practicalities mean that the 
sensitivity of the test is dependent on the spectral type of the star.
Li is depleted very rapidly on timescales of a few Myr in stars of type 
late K and M. This timescale rises to 100 Myr for early K stars and 
perhaps to 1 Gyr for F stars (see Soderblom et al. 1993). At the same 
time, the strength of the Li 6708 \AA\, resonance line, which is the only realistic 
diagnostic of Li abundance in stellar photospheres, declines drastically with increasing 
temperature at a given abundance. Thus, while the presence of undepleted 
Li is a necessary condition for F or G stars to be identified as PMS, 
it is not sufficient. For K and M stars the presence of undepleted Li is 
almost a guarantee of a PMS status. In practice this means we must 
identify a threshold EW for the Li 6708 \AA\, line that is consistent with an 
undepleted Li abundance (see Section 5.4).

\begin{figure}
\vspace{6.5cm}
\includegraphics{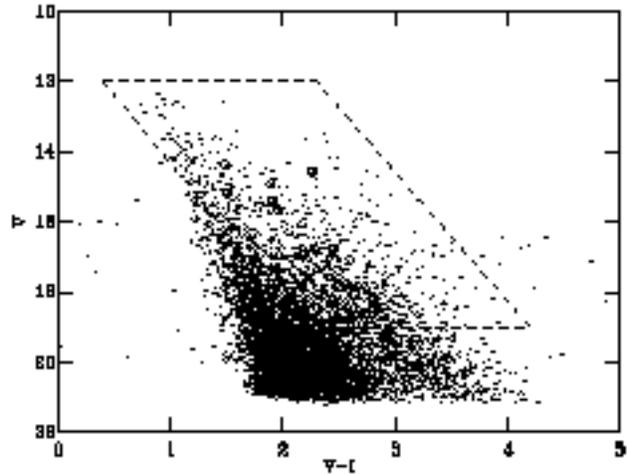}
\caption[The PMS objects selected from the PMS strip in V vs (V-I)]
{The PMS objects selected from the PMS strip (within the dashed box), encompassing the 
optical counterparts to the {\it ROSAT} X-ray sources of NF99 (circled).}
\label{selVI}
\end{figure}

The objects for the spectroscopic follow-up observations were selected
from the combined V versus (V-I) CMD for fields 1a, 2a, 3a,
4a, 2b, and 3b.
Fields 1b and 4b were excluded because they are pointing at the
molecular cloud, and are therefore most likely to contain only
foreground objects.
Had we wished to maximise the number of PMS stars in our spectroscopic
sample we would have chosen a region which fitted tightly around the
suspected (X-ray selected) PMS objects.
However, since our aim was to find the limits of the PMS in V vs
(V-I), we actually chose a region which encompasses the whole ``second
sequence'' and also the red side of the ``background'' sequence.
This sample should have little bias in terms of age and allow for
differential reddening across the Cep OB3b subgroup (see Fig.
\ref{selVI}).

\begin{figure}
\vspace{6.5cm}
\includegraphics{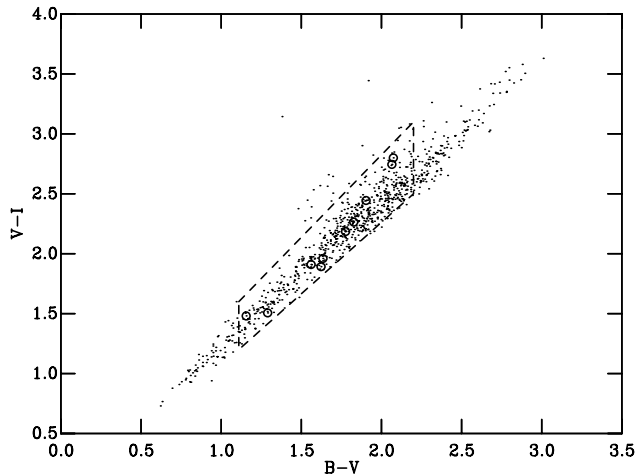}
\caption[Additional selection in (V-I) vs (B-V)]
{Additional selection from the possible PMS objects selected in the V versus (V-I) 
colour-magnitude diagram of Fig. \ref{selVI}. The resulting objects are contained 
in the dashed box, the optical counterparts to the {\it ROSAT} X-ray sources of NF99 are 
circled.}
\label{selBV}
\end{figure}

All the catalogue stars with good flags and photometric errors less
than 0.1 mag for V and (V-I), falling in the strip, were retained.  An
exception was made for those flagged as non-stellar objects, which were
retained since they could be double stars. We did not select stars
fainter than $V=19$ mag, mainly because they would be too faint for
spectroscopy. The result of this selection is shown in
Fig. \ref{selVI}: selected objects are within the dashed box, optical
counterparts to the {\it ROSAT} X-ray sources are circled. There are
1225 selected objects out of 7884.

We then performed a further selection 
by choosing stars having an error less than 0.1 mag and a good
flag in the (B-V) colour too, and picking up only those stars which
were contained in a strip in the (B-V) versus (V-I) colour-colour
diagram, again defined by the optical counterparts to the {\it ROSAT}
X-ray sources of NF99 (see Fig. \ref{selBV}).  The result of this
additional selection left 616 possible PMS objects.

These objects were the initial sample for the spectroscopic investigation
which was carried out with the multi-fibre spectrograph WYFFOS. 
The configuration program available for WYFFOS observations gave us two fields 
for a total of 110 (out of 616) optically-selected possible PMS objects.

\subsection{WYFFOS/AF2}

Intermediate-resolution spectra of the 110 optically-selected possible PMS objects 
were obtained on December 11-12, 1999, with the WYFFOS/AF2 multi-fibre spectrograph at 
the 4.2-m William Herschel Telescope (WHT), 
La Palma, Canary Islands, equipped with a CCD TEK 6 chip 
(type TK1024), the WYFFOS echelle grating (632 grating rulings/mm)
and the echelle order-sorting filter number 4 (6625 \AA) to cover the region occupied by the 
H$\alpha$ (6563 \AA) and Li\,{\sc i} (6707.8 \AA) features.

We obtained a wavelength coverage of about 440 \AA, in the approximate
range 6380-6820 \AA.  The nominal dispersion is 17.8 \AA/mm (equivalent
to 0.43 \AA/pixel), with a spectral resolution (FWHM) of about 2
pixels, equivalent to about 1 \AA, determined from arc lines. Each
fibre covers 2.7 arcsec in diameter on the sky.

Target, bias, arc, flat and offset sky frames were taken. Target frames
were observed for $3 \times 1800$s in the first configuration (first
night), and for $3 \times 1500$s in the second configuration (second
night).  Three offset sky frames (of 400s and 300s on the first and
second nights respectively), were taken for each configuration by
beam-switching the telescope to nearby sky (at 15 arcsec north, south
and west of the target position). These were taken in order to
normalise the fibre transmissions (throughput/vignetting correction),
under the assumption that the night sky has the same brightness through
each fibre. About 15 dedicated sky fibres were assigned in each
configuration and these were used to determine the mean sky spectrum.
In addition, we used the offset exposures to create a sky
spectrum for each fibre (which was a mean of the offset sky exposures
through that fibre).  Sky subtraction using the offset sky exposures
was used to create the spectra from which H$\alpha$ equivalent widths
(EWs) were measured, since this minimised the effects of nebular
contamination of the H$\alpha$ lines in target regions of high
nebulosity. The mean sky determined from dedicated sky fibres was
instead used to create the spectra which were cross-correlated and from
which Li\,{\sc i} EWs were measured, since this resulted in higher
signal-to-noise spectra.  In addition, on the second night we measured
some radial velocity standard stars through a single fibre at the
centre of the field.

The {\sc iraf}\footnote{{\sc iraf}(Image Reduction and Analysis Facility) is
distributed by the National Optical Astronomy Observatory (NOAO), which
is operated by the Association of Universities for Research in
Astronomy (AURA), Inc. under cooperative agreement with the National
Science Foundation.} {\sc wyfred} procedures (Pollacco et al. 2000) were
used to process the images and extract the multi-fibre spectra obtained
with WYFFOS. From the residuals to continuum fits in line-free regions
of the spectra, we derive a typical signal-to-noise of S/N$>10$ for the
majority of the targets.

\begin{figure}
\vspace{6.7cm}
\includegraphics{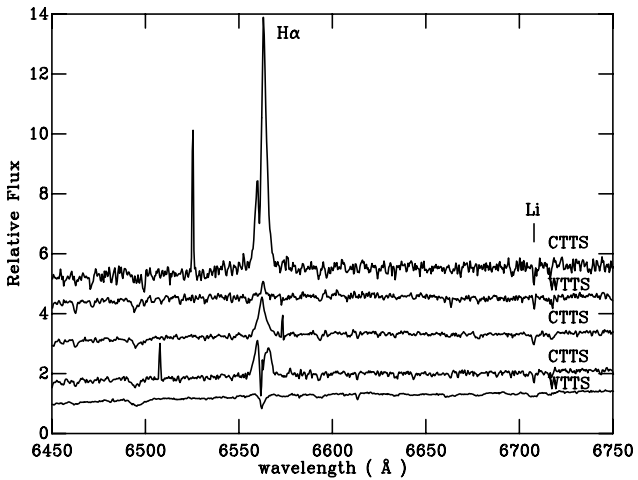}
\caption[The spectra (normalised to 1 and then offset) of the 5 TTS from the first 
configuration]
{The spectra of the 5 TTS from the first configuration. From bottom to top, respectively:
object 7 (field 2b), object 8 (field 2b), object 18 (field 1a),
object 32 (field 2a), and object 49 (field 3b).}
\label{spcep1_tot}
\end{figure}

\begin{figure}
\vspace{6.7cm}
\includegraphics{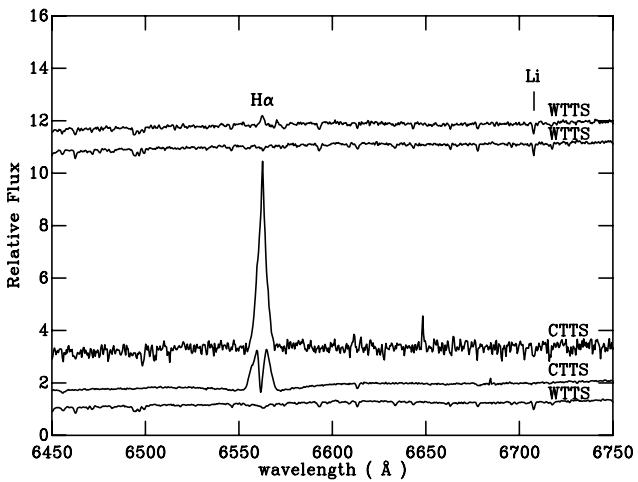}
\caption[The spectra (normalised to 1 and then offset) of the 5 TTS from the second 
configuration]
{The spectra of the 5 TTS from the second configuration. From bottom to top, respectively:
object 81 (field 3b), object 84 (field 3b), object 95 (field 1a), 
object 100 (field 2b), and object 106 (field 1a).
Note that object no. 84 is the veiled CTTS for which we do not have a RV measurement: 
the extension of the wings on both sides of the H$\alpha$ line suggests a strong T Tauri wind.} 
\label{spcep2_tot}
\end{figure}

\subsection{Radial velocities}\label{radveldet}

Radial velocity (RV) measurements were performed using the Starlink
software package {\sc figaro}. 
The spectrum of the IAU radial velocity standard star HD 213947
(spectral type K4 III) was extracted too. All the objects and standard
spectra were re-binned to a logarithmic wavelength scale, giving
constant velocity bins of 19.152 km s$^{-1}$.  The targets were
cross-correlated with the radial velocity standard star in the spectral
range 6400-6550 \AA\, (a region containing easily identifiable metallic
lines), in order to determine the relative shifts, and calculate RVs.
We also tried the region 6650-6800 \AA\, but the cross-correlation peak was
weaker and did not improve the result.  Note that possible cosmic rays
and N~{\small II} ($\lambda\lambda$ 6548, 6584) or S~{\small II}
($\lambda\lambda$ 6717, 6731) nebular emission lines were masked before
performing the cross-correlation.  In the case of double peaks in the
cross-correlation function, the cross-correlation was repeated in the
region 6650-6800 \AA: if both peaks were still separated (indicating a
possible binary nature of the object), we fit a double Gaussian to find
the shifts (and hence RVs) for both of them. In total we were able to
obtain RV measurements for 74 objects. Of the 36 targets without RV
determinations, 15 are earlier type A and F stars, while the rest have
relatively noisy spectra.

From repeat measurements of a number of targets taken for a different
programme on the same observing run, we estimate a RV accuracy of $\pm
3$ km s$^{-1}$.  The relative RVs were placed in the heliocentric
reference frame.

Radial velocity measurements allow us to find candidate kinematic
members of the Cep OB3 association. We consider to be likely members
all those stars with RV values within 2 sigma ($=6.0$ km s$^{-1}$) of
the Cep OB3 mean RV value $-22.5 \pm 1.5$ km s$^{-1}$ (derived from
already known members; Garmany 1973), i.e., in the range $-28.5$ to
$-16.5$ km s$^{-1}$.  Out of 74 objects with measured RV, we found 21
stars in total as candidate kinematic members and 53 kinematic
non-members.  We can estimate the likely number of contaminants to have
the correct RV by chance.  Assuming for the 53 kinematic non-members a
uniform spread over about 190 km s$^{-1}$ velocity space, then
we would only expect about $53*12/190 = 3.3 \pm 1.8 $
objects to fall by chance in our selection range of $\pm 6$ km/s.
Therefore, out of 21 candidate kinematic members, we would expect about
$3 \pm 2$ objects to have the correct RV by chance and $18 \pm 2$
objects to be true association members.

\subsection{Li\,{\sc i} and H$\alpha$ equivalent widths}

The Li\,{\sc i} ($\lambda 6707.8$) and H$\alpha$ ($\lambda 6563$) EWs
were measured using the Starlink {\sc dipso} package.  The EW was measured by
integration below a linear continuum between the two extremes points of
the specified line. Each EW was measured three times (because of
possible uncertainties in the continuum placement), and a mean value
determined.  The mean estimated errors are of about 20 m\AA\, and
0.2\AA\, respectively for Li\,{\sc i} and H$\alpha$ EWs, as a result of
uncertainty in the continuum location.

Note that at our resolution the Li\,{\sc i} 6708 \AA\, line cannot be
separated from an Fe\,{\sc i} line at 6707.4 \AA, although strong
Fe\,{\sc i} lines at 6705 \AA, and 6710 \AA\, are clearly
resolved. However, the likely strength of the 6707.4 line for F- to K-type
stars would be $\leq$ 25 m\AA\ -- its EW being estimated from the
empirical relationship [20(B-V)$_0$ $- 3$] m\AA\, given by Soderblom et
al. (1993).

By comparing the spectra with the library of stellar spectra given by
Jacoby, Hunter \& Christian (1984), we can also give a rough spectral
classification for the objects.  In Figs. \ref{spcep1_tot} and
\ref{spcep2_tot}, a montage of some of the extracted spectra is
presented.  The objects shown are all PMS stars according to our
classification method (see below).

\begin{figure}
\vspace{6.7cm}
\includegraphics{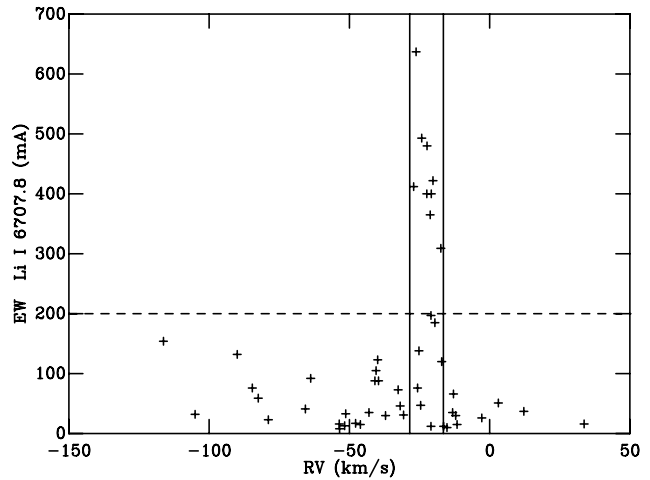}
\caption[.....]
{The 49 objects for which we have a RV measurement and which show Li\,{\sc i} 6708 
in absorption. The vertical lines define the RV range appropriate to members of the
Cep OB3b subgroup. Objects above the dashed horizontal line at 200 m\AA\, are definite
K-type PMS stars. Those below are all F and G type stars, apart from one K-type star
with a  Li\,{\sc i} EW of 197 m\AA. Note that objects with only upper limits for Li\,{\sc i} 
6708 absorption are not shown.}
\label{LiRV}
\end{figure}

In Fig. \ref{LiRV}, we plot the 49 stars for which we have both
measured Li\,{\sc i} 6707.8 \AA\ EWs and RVs, marking out the range for
kinematic membership specified in Section \ref{radveldet}.  For the
other 25 stars with measured RV we cannot detect an Li\,{\sc i} feature
(5 of these have RVs qualifying them for membership of Cep OB3b). Data
on the 21 objects that have a RV consistent with membership of Cep OB3b 
are listed in Table 4.
We give their coordinates, RVs, photometric values, Li and H$\alpha$
EWs, spectral types, class and correlation with a {\it ROSAT} X-ray
source of NF99.

For those stars with measurable Li\,{\sc i} absorption, we can set a
minimum EW above which classification as a definite PMS object can be
considered. A limit of 200 m\AA\ is chosen for K stars, following the scheme
proposed by Mart\'{\i}n (1997). This is the EW expected for a K0 star
just leaving the PMS phase of evolution. For M stars the EW associated
with this transition is about 500 m\AA.  We find 9 kinematic members
with a Li\,{\sc i} EW $>200$ m\AA: since these are all K-type stars, we
can accept them as Cep OB3b PMS members.  Data on the 12 remaining
kinematic members of the association are given in the lower part of
Table 4. Two of these stars (no. 13 and 69) are K stars with
insufficient Li EW to be considered as young PMS objects (note however
that no. 13 only misses the 200 m\AA\ limit by 3 m\AA\, so we consider
it possible PMS as well).  The remaining 10 are F- or G-type
stars. There are 6 of them which are ``possible'' PMS members, due
to their spectral type and Li\,{\sc i} absorption.  For spectral types
earlier than K the empirically determined timescale for Li depletion
becomes long enough that an EW indicating essentially undepleted Li is
no longer a guarantee of PMS status, but merely means the star is young
(say less than a few hundred Myr).  However if the Li EW is
significantly less than expected for undepleted photospheric Li, then a
PMS status can be ruled out. The appropriate EW thresholds can be
judged from measurements in young open clusters like the Pleiades
(Soderblom et al. 1993) and are 100-200 m\AA\, for G-type stars,
50-100 m\AA\, for F-type stars and perhaps even no detection of the Li
line at all for stars of type F0 or earlier.  Note that we do not
consider object no. 50 in Table 4 as a possible PMS star: it is F-type
with a Li EW which could just be consistent with undepleted Li; its
$(V-I)_0$ is about 0.35 mag and thus $E(V-I)=2.5$, hence implying an
$A_V$ value of $\sim 6$ which is much higher than any other extinction
value reported for known Cep OB3b members.

In total we find 16 objects which are likely to be Cep OB3b PMS members
- reasonably consistent with our expectations from the RV measurements (see Section
\ref{radveldet}) which predicted about $18 \pm 2$ objects to be true
association members.

\begin{figure}
\vspace{6.7cm}
\includegraphics{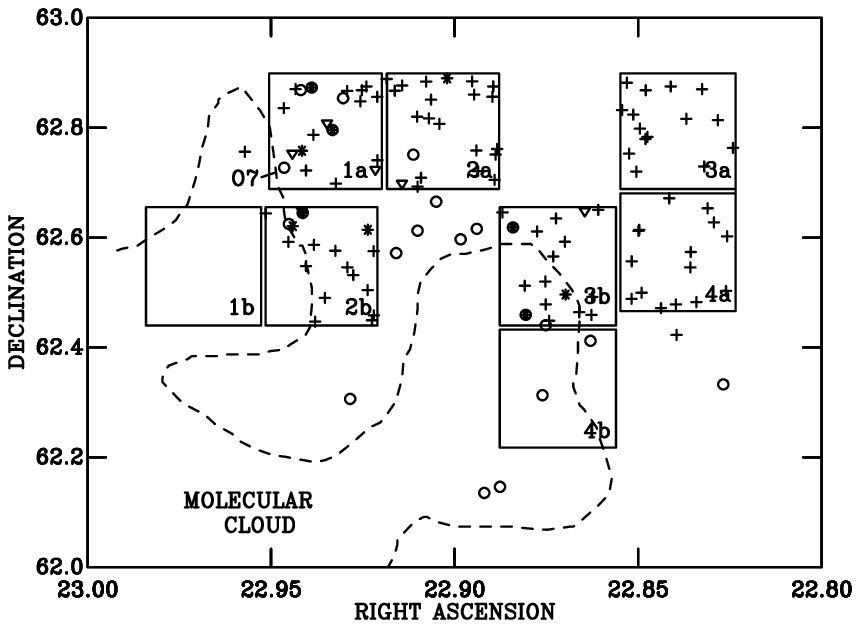} 
\caption[]
{The spatial distribution of the 16 PMS objects in the Cep OB3b subgroup (symbols as in Fig. 
\ref{figfields3zoom}). The 5 CTTS and 5 WTTS are shown as filled circles and asterisks 
respectively. The 6 possible PMS stars are shown as open triangles. Note that one of the 
possible PMS stars is hidden behind the CTTS in field 2b. The remaining 94 spectroscopic
objects are shown as black crosses (note that two of them fall outside our CCD fields:
they are stars which fell down a sky fibre).}
\label{figfields3zoom_TTS}
\end{figure}

\begin{table*}
 \centering
 \begin{minipage}{170mm}
  \caption{The candidate PMS and possible PMS members of Cep OB3b found with the present 
investigation.}
\begin{tabular}{@{}rcllcccccccc@{}}
 $\#$ & field & RA (J2000.0)   &  Dec (J2000.0) & RV$_{hel}$     & V      & (V-I)      & EW(Li\,{\sc
 i})  & EW(H$\alpha$) & sp. type & class & X-ray \\
  &  &  &     & (km s$^{-1}$)    &        &            &    (m\AA)        &     (\AA) & & & (NF99)  
\\[10pt]
  7\footnote{This object is cross-correlated with BHJ40-096 (J96). According to NF99, it is 
K-type and has EW(Li\,{\sc i})$=380$ m\AA \, and EW(H$\alpha$)$=0.6$ \AA, whereas we find values 
about 30 and 80 per cent larger.} & 2b & 22 56 38.721 & +62 37 14.43 & -24.2 & 14.548 & 2.264 & 493 &  1.1 & K & WTTS & X34  \\
  8 & 2b & 22 56 28.955 & +62 38 41.49 & -20.2 & 16.733 & 2.347 & 422 & -5.9 & K & CTTS &  \\
 18 & 1a & 22 56 20.132 & +62 52 23.16 & -22.3 & 16.789 & 2.444 & 480 & -6.2 & K & CTTS & X26 \\
 32 & 2a & 22 54 07.285 & +62 53 42.23 & -27.1 & 16.939 & 2.113 & 412 & -0.9 & K & WTTS &  \\
 49\footnote{This is a PMS object (CTTS), but, unfortunately, it is on a 
bad column on the CCD image and so it has a non-zero flag (see Naylor et 
al. 2002), therefore we cannot trust its V ($=16.030$) and (V-I) 
($=1.867$) photometric colours.}  
    & 3b & 22 53 03.636 & +62 37 08.27 & -20.8 &  (b)   &  (b)  & 400 & -27.3 & K & CTTS &  \\
 81 & 3b & 22 52 11.102 & +62 29 46.71 & -17.4 & 15.676 & 2.113 & 309 &   0.5 & K & WTTS &  \\
 84\footnote{We also consider this object to be a kinematic member, even 
though we do not have its RV measurement, because of its CTTS nature; and it is clearly veiled.} 
    & 3b & 22 52 50.152 & +62 27 33.83 &  :    & 15.626 & 2.013 & :   & -13.4 & K & CTTS &  \\
 95 & 1a & 22 55 59.654 & +62 47 45.08 & -26.2 & 17.128 & 2.626 & 637 & -28.4 & K & CTTS & X20 \\
100 & 2b & 22 55 25.087 & +62 36 51.35 & -21.2 & 16.180 & 2.028 & 365 &   0.3 & K & WTTS &  \\
106 & 1a & 22 56 29.678 & +62 45 28.68 & -22.4 & 15.660 & 1.960 & 400 &  -0.8 & K & WTTS & X29 \\
 & & & & & & & & & & &  \\
 13 & 1a & 22 55 17.649 & +62 43 23.43 & -20.9 & 15.783 & 1.593 & 197 &   0.8 & K & PMS? &  \\
 16 & 1a & 22 56 05.055 & +62 48 27.26 & -25.2 & 15.912 & 1.543 & 138 &   2.4 & G & PMS? &  \\
 36 & 3b & 22 51 52.222 & +62 38 51.49 & -19.5 & 15.670 & 2.108 & 185 &   1.4 & G & PMS? &  \\
101 & 2b & 22 56 29.791 & +62 38 53.76 & -17.1 & 15.727 & 2.395 & 120 &   2.1 & F & PMS? &  \\
 & & & & & & & & & & &  \\
 10 & 2a & 22 54 51.609 & +62 41 52.40 & -25.7 & 15.004 & 1.962 &  76 &   4.2 & F & PMS? &  \\
 56 & 3a & 22 51 04.777 & +62 49 25.35 & -20.9 & 16.853 & 2.180 &  12 &   2.3 & G & &  \\
 69 & 4a & 22 50 58.865 & +62 36 50.63 & -24.6 & 15.553 & 1.511 &  47 &   1.0 & K & &  \\
 & & & & & & & & & & &  \\
 12 & 1a & 22 56 39.087 & +62 45 09.36 & -20.3 & 14.910 & 1.892 & $<10$ & 3.6 & F & PMS? & X35 \\
 38 & 3b & 22 52 20.510 & +62 38 05.39 & -27.0 & 14.770 & 1.259 & $<10$ & 1.3 & G & &  \\
 40 & 3b & 22 50 52.516 & +62 52 05.88 & -23.7 & 16.920 & 2.116 & $<10$ & 1.2 & G & &  \\
 50 & 3b & 22 51 45.635 & +62 27 32.94 & -27.3 & 14.676 & 2.841 & $<10$ & 3.6 & F & &  \\
 76\footnote{This is a star falling down a sky fibre, for which we do not have optical photometry.} 
    &    & 22 50 22.26 & +62 25 21.6 &  -24.1 &       &       & $<10$ &  1.7 & G & & \\
\end{tabular}
\end{minipage}
\end{table*}

\begin{table}
 \centering
 \begin{minipage}{83mm}
  \caption{Profiles and velocity width of the H$\alpha$ emission line for the PMS members.} 
  \begin{tabular}{@{}cccc@{}}
 $\#$ & class & type \footnote{From Reipurth et al. (1996) scheme: I for symmetric profiles with no or weak absorption features; II for double peak profiles, with the secondary peak more than half the strength 
of the primary one; III for double peak profiles with the secondary peak less than half 
the strength of the primary peak. B and R define the location of the secondary peak 
with respect to the primary (i.e., blue-wards or red-wards).} & $V_{H\alpha}$ (km s$^{-1}$) \\
      &       &      &    (FWZH)\footnote{FWZH = full-width at zero height} \\[10pt]
  7 & WTTS &       & 460 \\
  8 & CTTS & II-R  & 630 \\
 18 & CTTS & I     & 630 \\
 32 & WTTS & III-B & 260 \\
 49 & CTTS & III-B & 620 \\
 81 & WTTS &       & 430 \\
 84 & CTTS & II-B  & 830 \\
 95 & CTTS & I     & 670 \\
100 & WTTS &       & 410 \\
106 & WTTS & I     & 260 \\
\end{tabular}
\end{minipage}
\end{table}

The EW of the H$\alpha$ line can be used to further classify a PMS
object as a CTTS or WTTS.  In fact, although its Li\,{\sc i} EW can be small
because of optical veiling (responsible for a masking of the absorption
lines; see for example Basri, Mart\'{\i}n \& Bertout 1991), a CTTS can
be immediately recognised by its very wide and strong H$\alpha$
emission which cannot be explained by means of stellar chromospheric
activity only (e.g., T Tauri winds, active accretion discs, or magnetic
accretion columns; see Appenzeller \& Mundt 1989; Hartmann, Hewett \&
Calvet 1994).  Therefore it must show an emission H$\alpha$ EW larger
than that expected from chromospheric emission for its spectral
type. Conservative EW emission values proposed by Mart\'{\i}n (1997; see also
references therein) are: greater than $5$ \AA \, for K-type
stars, $10$ \AA \, for early M-type stars, and $20$ \AA \, for late
M-type stars.  On this basis, we can classify 4 of the 9 definite PMS candidates
(with Li\,{\sc i} EW $> 200$ m\AA) as CTTS, the other 5 PMS stars
are WTTS. None of the remaining 6 ``possible'' PMS stars show strong
H$\alpha$ emission.

Note that another K star (no. 84 in Table 4), without a useful RV and
Li\,{\sc i} EW measurement because it is so heavily veiled, can be
classified as CTTS in Cep OB3b as well, by virtue of its extremely
strong H$\alpha$ emission.

The spectra of these 5 CTTS and 5 WTTS are shown in
Figs. \ref{spcep1_tot} and \ref{spcep2_tot}.

We checked the velocity width of the H$\alpha$ emission lines in the 10
TTS and found that they were much larger than any plausible amount of
rotational broadening (see Table 5; the typical error is about $9$ km
s$^{-1}$).  From the strength of the line in each of them (full-width
at zero height) we can say that possible problems in the sky
subtraction procedure are not going to change the results of our
classification (any sky line contamination is unresolved at FWHM $\sim$
50 km s$^{-1}$).  In general, the spectroscopic targets having
H$\alpha$ in emission show single-peaked as well as more complex
profiles.  A classification in these terms is also given in Table 5,
using the scheme established by Reipurth, Pedrosa \& Lago (1996).

We note that out of 10 definite TTS, just 4 (2 WTTS and 2 CTTS) have a
{\it ROSAT} X-ray counterpart in NF99. Although the NF99 survey was not
very sensitive, the low fraction of detections is an example
of how an X-ray selected PMS sample can be incomplete compared to
optically-selected samples, unless the original X-ray survey is deep
enough to sample the entire X-ray luminosity function of PMS stars in the
considered mass range.

Out of 74 targets with measured RVs, there are 9 TTS plus a further 6
possible PMS stars which are Cep OB3b kinematic members and 53
non-members: this suggests that the contamination in the optically-selected 
PMS sample is of order 70 per cent.  However, this large
percentage is most likely the consequence of the generous region over
which the targets were initially selected in the V versus V-I CMD.

The location of the CTTS, WTTS and possible PMS stars with respect to
known Cep OB3b members is given in Fig. \ref{figfields3zoom_TTS}. 
Also plotted there are the remaining 94 spectroscopic objects.
This confirms that low-mass PMS stars may indeed form close to
higher-mass companions. The confirmed TTS seem to be spread in all the fields
other than 3a and 4a, which are the most distant from the CO contour of
the molecular cloud.  Note, however, that these are small-number
statistics and that we are biased against the detection of TTS in
fields 3a and 4a simply by the field of view and spatial distribution
of allocated fibres in the WYFFOS configurations.

By means of the spectroscopic identification of at least 9 T-Tauri stars with
the appropriate RVs, we have confirmed the presence of low-mass stars
associated with the younger subgroup of the Cep OB3 association.

\begin{figure}
\vspace{8.0cm}
\includegraphics{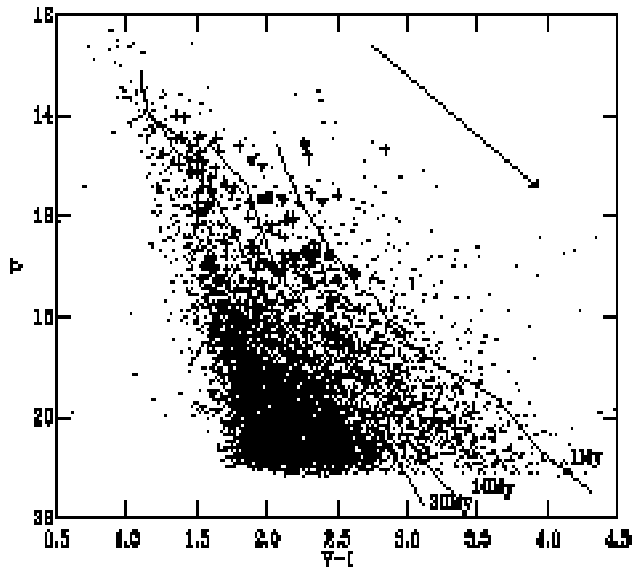} 
\caption[]
{The colour-magnitude diagram in V versus (V-I) for the stars in the six fields used
for the selection of the PMS sample for the spectroscopic follow-up. 
CTTS and WTTS from the spectroscopic follow-up are shown as filled circles and asterisks 
respectively.
The remaining Cep OB3b kinematic members which are possible PMS stars are shown as 
open triangles. The other objects which were spectroscopically observed are shown as black 
crosses (two stars which fell down a sky fibre and for which we do not have photometric 
data are not shown). 
Note that a possible PMS object is hidden behind the WTTS at V $= 15.7$ mag
and (V-I) $= 2.1$ mag; and one CTTS is not shown because we do not trust its 
photometric colours. Also plotted there are isochrones for 1, 10 and 30 Myr 
(Siess et al. 2000) and the average reddening vector for Cep OB3b (see Section
\ref{cmdsec}).}
\label{plotvviTTS}
\end{figure}

\section{The age of the TTS objects}\label{isofit}

How old are the PMS objects in Cep OB3b? Is there an age gradient across the Cep
OB3b subgroup?  To answer these questions we have to compared
the V versus (V-I) CMDs for different fields by means
of isochrone fitting.  If one can fit the PMS stars in one field with
different isochrones, the most tempting conclusion would be a spread in
ages.  However, as we discuss below, there are several other factors which
could contribute to such a spread.

\begin{figure*}
\vspace{14.0cm}
\includegraphics{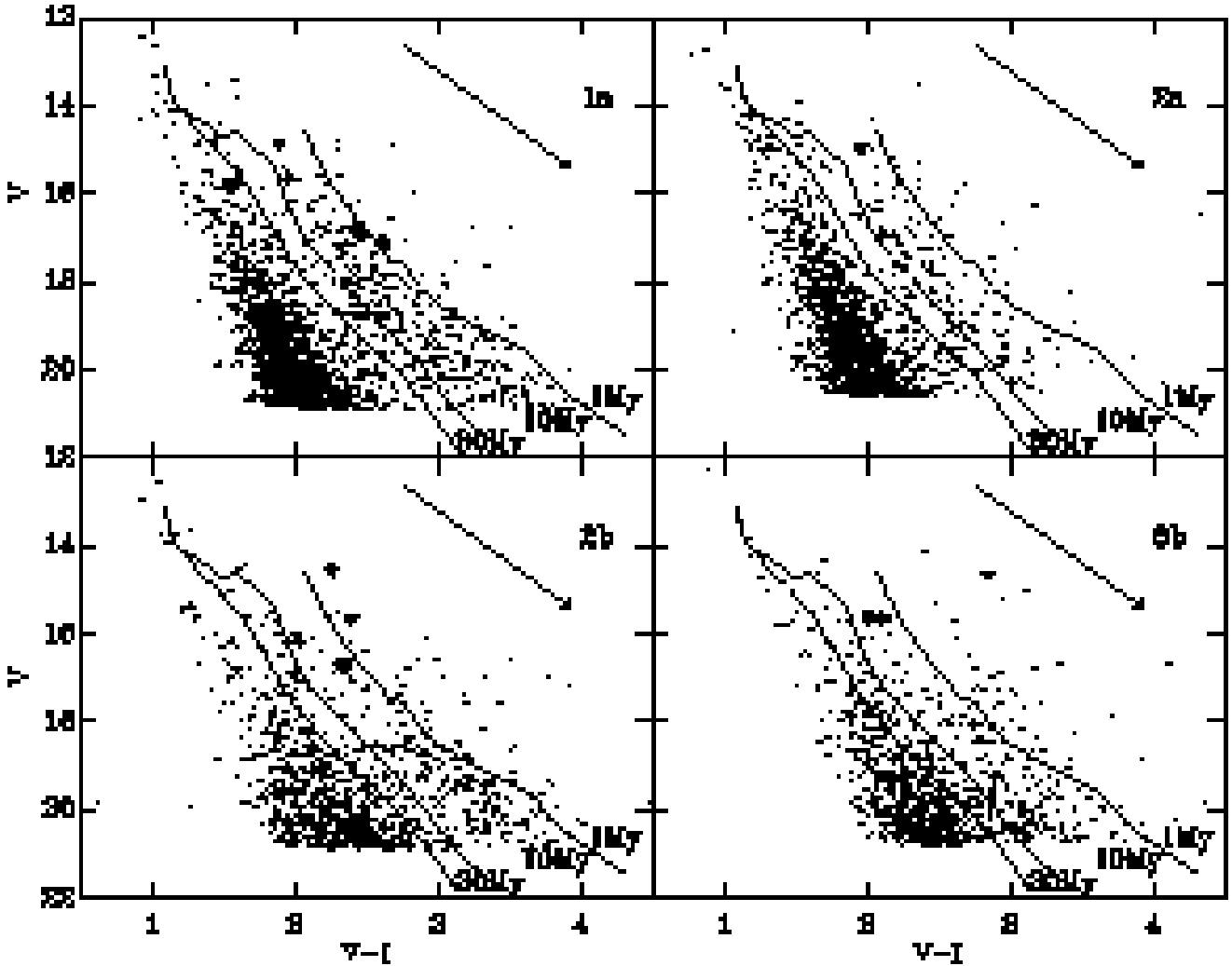}
\caption[Isochrones for the field 1a and field 2a]
{Isochrones for 1, 10 and 30 Myr (Siess et al. 2000) for fields 1a (top-left), 2a (top-right),
2b (bottom-left) and 3b (bottom-right).
CTTS and WTTS are shown respectively as filled circles and asterisks. 
The remaining Cep OB3b kinematic members which are possible PMS stars are enclosed by
open triangles. 
Note that 1 CTTS belonging to field 3b is not plotted because we do not trust its 
photometric colours (see footnote b in Table 5). Furthermore, note that a possible PMS 
object in field 3b is hidden behind the WTTS. And finally, always in field 3b, the open 
triangle which is not centred on a data-point is a possible PMS object 
which is flagged in our catalogue (and therefore not appearing in the plot), but which was 
included in the sample for the spectroscopic-follow up due to its possible binary nature: 
it is a close visual binary on the CCD image (and indeed a close examination of the 
cross-section of the CCD image revealed that it is double peaked), but it is not a 
spectroscopic binary. Also plotted there is the average reddening vector for Cep OB3b 
(see Section \ref{cmdsec}).}
\label{plotisochs_TTS}
\end{figure*}

Isochrones were computed from the evolutionary models of Siess, Dufour
\& Forestini (2000), using a metallicity of Z=0.02 and masses in the
range $0.2 < M/M_{\odot} < 3.0$.  To convert the effective temperatures
resulting from their models to colours, we chose to use the conversion
table of Siess, Forestini \& Dougados (1997).  In Figs. \ref{plotvviTTS}
and \ref{plotisochs_TTS} we plot isochrones corresponding to 1, 10 and
30 Myr, applying the mean extinction ($A_V=2.81$), colour excess
($E(V-I)=1.18$) and distance modulus ($dm=9.65$) for Cep OB3b (see
Section \ref{cmdsec}).

Figure \ref{plotvviTTS} shows the location of both the TTS and the
possible PMS Cep OB3b kinematic members in the V versus (V-I)
colour-magnitude diagram. Also shown there are the remaining objects which
were spectroscopically followed-up (apart from two stars which fell down a 
sky fibre and for which we do not have photometric data). 
It can be seen in this diagram that the 6 possible PMS stars are more widely scattered 
in (V-I) than the TTS, with a few sitting in the locus of the background sequence. 
At the same time, the V magnitudes for this group occupy a similar range to the TTS
despite the group's much earlier mean spectral type. Given the
uncertainty over the status of this group, we restrict further
discussion to the definite TTS only.

In field 1a (see Fig. \ref{plotisochs_TTS} [top-left]) the TTS objects
are likely to be 1-10 Myr old, in agreement with the age ($\sim 7$ Myr)
determined for the OB-type members. More precisely, the 2 CTTS (shown
as filled circles) lie close to the 1 Myr isochrone, whereas the WTTS
(shown as an asterisk) lies closer to 10 Myr.  Note that there is a
lack of stars between the 10-30 Myr isochrones: this clearly defines
the PMS locus in the CMD -- the bulk of stars occupying the region
between the 1-10 Myr isochrones with a few at even younger ages.  We
have repeated the analysis using empirically calibrated isochrones from
the D'Antona \& Mazzitelli (1997) models, which yield slightly younger
ages.

In Field 2a (see Fig. \ref{plotisochs_TTS} [top-right]) we have just 1
WTTS, just less than 10 Myr old, suggesting an age similar to that of the
WTTS in field 1a. Again, the interval between the 1-10 Myr isochrones is
more populated than the interval between the 10-30 Myr isochrones.

In field 2b (see Fig. \ref{plotisochs_TTS} [bottom-left]) we have 1 CTTS
showing an age of about 1-5 Myr and 1 WTTS closer to 10 Myr old, in
agreement with those found in field 1a; but there is also a WTTS that
appears to be less than 1 Myr old.  This WTTS could be as young as
this, but a $0.8$ mag difference in its visual magnitude could also perhaps
be explained as due to binarity, photometric errors and
spread in distance (discussed below). Variations in extinction or the
value of $R$ are unlikely to be of assistance. The reddening vector is
close to parallel to the isochrones, whilst an anomalously large value
of $R$ would steepen the reddening vector and lead to an even younger
age (see Section 3). If we look at its position with
respect to the Cep OB3b subgroup (see Fig.  \ref{figfields3zoom_TTS}
and Table 5), we can see that this object is closer to the molecular
contour and to the early-type member BHJ 40: although a larger $A_V$
value would be plausible because of its proximity to the molecular
material, this seems not to be the case given that BHJ 40 has an
extinction of just $\sim 3$ mag.  
In field 2b there seems to be no clear difference in
the stellar densities of the gaps between the 1-10 Myr and the 10-30
Myr isochrones: we think that the 10-30 Myr gap is filled by some
objects of the background population suffering higher extinction,
since the bulk of the background sequence in this field appear to lie
at redder (V-I) colours than that in field 1a.  In Appendix B we report
the additional extinction necessary to match the background sequence in
field 2b with that of field 1a.

Finally, in field 3b (see Fig. \ref{plotisochs_TTS} [bottom-right]), there
are 1 CTTS and 1 WTTS at about the same age of
$\simeq 5$\,Myr. Note that 1 CTTS belonging to this field is not plotted because
we do not trust its photometric colours (see footnote b in Table 5).
These two TTS are even closer to the cloud than the TTS of field 2b
(see Fig.  \ref{figfields3zoom_TTS} and Table 5). However, an anomalous
R value for these two objects, which would give an even younger age, is
once again unlikely: the extinction for BHJ 10 (an early-type member
falling in this field) is $A_{V} \simeq 2.5$, a little less than the
adopted extinction. 
As in field 2b, there is no clear gap between the 10-30 Myr isochrones:
this effect is again probably caused by contamination from the
background sequence, pushed to even redder colours than in field 2b
(see Appendix B).  Furthermore, there seem to be somewhat more objects
in this field which lie on the red side of the 1 Myr isochrone.

It is not easy to assess the vertical spread around the isochrones in
each of the fields 1a, 2a, 2b and 3b respectively. In fields 2b and 3b
the contamination from background objects complicates the picture, and
it also seems that there are many objects at even younger ages ($< 1$
Myr).  These objects could be early-type PMS stars very deeply embedded in the
molecular cloud than the TTS objects we have discovered, i.e.,
suffering an higher extinction than the mean adopted value of 2.81, but
the extinction vector runs nearly parallel to the isochrones, so this
seems unlikely (see below). A higher value of R would make these
objects even younger. Neither are these stars likely to be
foreground contaminants. The CMDs of fields 1b and 4b which are
pointing at the molecular cloud do not show many stars in this portion
of the colour-magnitude diagram.  We suspect therefore that there are
genuine, very young low-mass PMS stars in Cep OB3b.
However, from Fig. \ref{plotvviTTS}
we can see that this possible $<1$\,Myr population has not yet been
adequately probed with spectroscopy, whereas we {\it can} rule out a
significant population of $>10$ Myr objects, because we have sampled
this area of the CMD quite well.

If we consider fields 1a and 2a, the separation between background and
PMS loci are clearly defined.  The photometric data alone suggest a
spread of ages between $<1$ and 10 Myr, but this is not contradicted by
the position in the CMDs of the small number of spectroscopically confirmed TTS
(see Fig. \ref{plotisochs_TTS}, top).  This corresponds to a vertical
spread of 1.5 magnitudes in V.  This is more than the spread expected
from the presence of unresolved binaries (0.75 mag) plus our photometric errors
for V (less than a hundredth of a mag up to V$=17.5$, and $\sim 0.05$ 
up to V$=20.0$). If we also allow for a spread in distance,
which is at most 0.1 mag for the Cep OB3b members within our fields, we
could perhaps explain a 0.9 mag spread.

It would then be tempting to explain the remaining 0.6 mag difference
as the effect of extinction variations. However, we stress again that
the reddening vector runs almost parallel to the 1 Myr isochrone and
therefore it is difficult to explain such a difference with a change in
the extinction value. A scatter of R between values of 3 and 5 would
introduce a horizontal scatter of about 0.2 in $V-I$ and hence about
0.5 in $V$, but there really is no concrete evidence that such a
scatter is present from the spectroscopic work on the early-type
association members.  Note that Preibisch \& Zinnecker (1999) found a
similar PMS CMD for stars in the Upper Sco OB association, with a
spread of about 1.2 mag, which they explained as the effect of
unresolved binaries, photometric errors and spread in distance.  They
concluded by saying that a small spread in ages cannot be excluded, but
that it is probably not larger than about 2 Myr. Dolan \& Mathieu
(2001) found instead a real age spread in the $\lambda$ Orionis PMS
population. Similarly, our results strongly suggest the possibility of
a real spread in ages for the low-mass stars in the Cep OB3b
subgroup. However, a larger spectroscopic sample is really needed to
confirm this hypothesis.

\section{Discussion}\label{secdiscus}

We have photometrically surveyed the younger subgroup of the Cep OB3
association, Cep OB3b, at optical wavelengths (see Sections
\ref{datared} and \ref{cmdsec}) and followed-up spectroscopically an
optically-selected PMS sample (see Section \ref{specclass}).  As a
result, we have discovered 10 TTS kinematic members of the association
and a further 6 objects that are very likely association members.
Isochrone fitting (see Section \ref{isofit}) to these objects suggests
ages from $<1$ to nearly 10 Myr, probably compatible with the
OB-type members of the subgroup. Furthermore, we have found that
binarity, photometric errors, spreads in distance and in extinction (as
suggested by the early-type members of the younger subgroup) are unlikely
to fully explain the spread of objects around the 1 Myr isochrone in the
CMDs. Unless there are large variations in $R$ or visual extinction,
for which there is no evidence from the early-type stars in the
region, then we have a quite strong suggestion of a significant age
spread in Cep OB3b.  Unfortunately, our spectroscopic sample is
small and in particular does not adequately sample the photometric
association candidates with apparent ages of $<1$\,Myr.

In the Introduction we anticipated speculations about how star
formation is proceeding in the Cep OB3 association, i.e., following the
sequential model of Elmegreen \& Lada (1977) or as the result of a
supernova explosion.  The sequential model would predict: bursts of
star formation occurring every few Myr, about 10-50 pc apart; a
substantial amount of gas near the younger subgroup, but none (or very
little) in the older subgroup; a velocity difference between the two
subgroups (or gas velocity, or both) in the range $5-10$ km s$^{-1}$.
This is effectively what is occurring in the Cep OB3 association, with:
a temporal separation between the older and the younger subgroup of
about 3 Myr and a spatial separation of about 13 pc (Sargent 1979); the
younger subgroup is closer to the molecular cloud (Blaauw 1964); and
there is a velocity difference of about $4$ km s$^{-1}$ (Garmany 1973)
between the two subgroups.
Unfortunately our spectroscopic sample is too sparse to be useful for
a detailed comparison. 
All we can say is that the confirmed PMS objects in fields 1a, 2a, 2b, and 3b
are between 1 and 10 Myr old.
In addition, field 3b shows more young objects less than 1 Myr old, potentially
implying that star formation in field 3b ended later than in field 1a.

Thus the photometric evidence leads us to marginally favour a
``contemporaneous'' model of star formation in the younger subgroup of
the Cep OB3 association, i.e., of star formation possibly triggered by
a supernova explosion. As suggested in the Introduction, the pulsar PSR
2324+60 of Assousa et al. (1977), some 10 Myr old, seems to be the
possible stellar remnant of a SN event, as testified by the presence of
a H~{\small I} expanding shell centred in the older subgroup. Its
location close to the easternmost fields could also explain why star
formation in the westernmost fields ended later, although another
possibility for terminating star formation is discussed below. 
Certainly the easternmost fields are closer to the major source of ionising radiation
in the younger subgroup, i.e., the O7n star: a more plausible
explanation for the likely age spread would then be that this star may
have made star formation more difficult in the recent past by
dispersing the shielding molecular material from its neighbourhood.

In total, we have found 10 TTS in the Cep OB3b subgroup, divided into 5
CTTS and 5 WTTS.  From the numbers of WTTS and CTTS, we can derive the
WTTS/CTTS ratio, a very important number in addressing the problem of
the dissipation timescales for circumstellar discs.  The value for the
WTTS/CTTS ratio presented in the literature changes significantly in
different star-forming regions. In T associations it is in
the range 1-13.  In particular, WTTS/CTTS values from 1 (central
region) to $>8$ (wider region), 2-8, 4 and 13 were found respectively
in Taurus-Auriga (Neuh{\"a}user et al. 1995; Hartmann et al. 1991),
Chamaeleon (Feigelson et al. 1993; Alcal\'a et al. 1995), $\rho$
Ophiuci (Mart\'{\i}n et al. 1998) and Lupus (Krautter et al. 1997).
All these studies are based on an optical spectroscopic follow-up of
X-ray selected samples, except for the study of Hartmann et al. (1991)
which is based on a proper motion selected sample. Note however that
these {\it Einstein} or {\it ROSAT} 
X-ray selected samples are biased towards WTTS, because CTTS are more
difficult to detect in soft X-rays, and therefore the true WTTS/CTTS
ratio may be even smaller.

There are not so many determinations of the ratio between TTS subtypes
in OB associations.  To our knowledge, the only two for which the
fraction of WTTS on CTTS has been determined (on the basis of the
strength of their H$\alpha$ emission lines) are the Orion and
Upper-Scorpius OB associations.  In these regions, the low-mass PMS
populations have about the same age ($\sim 6-7$ Myr and $\sim 5$ Myr
respectively). From an optically-selected sample followed by optical
spectroscopy, Dolan \& Mathieu (1999) found a WTTS/CTTS ratio of 17 in
$\lambda$-Ori.  Recently, Dolan \& Mathieu (2001) found a WTTS/CTTS
ratio of 35 in the central region of $\lambda$-Ori, from an optical
photometric survey followed up by multi-object spectroscopy.  In
Upp-Sco, Walter et al. (1994) found a WTTS/CTTS ratio of 14 from an
Einstein Observatory X-ray selected sample of PMS stars, together with
optical spectroscopic and optical and NIR photometric follow-up.
Preibisch \& Zinnecker (1999) found a WTTS/CTTS value of 24 (just 4
CTTS retaining their discs against 94 WTTS) from an X-ray selected
sample ({\it ROSAT} All Sky Survey and Einstein Observatory) of stars in the
same region, followed by optical spectroscopy and photometry; a 2dF
optical spectroscopic follow-up has recently revealed 98 new low-mass
PMS stars in a 6 square degree field in the association, with a
WTTS/CTTS of 9 (Preibisch et al. 2001).

The WTTS/CTTS ratio of 1 we have found for Cep OB3b is closer to the
value known for T associations than OB associations.  Furthermore,
although our sample is small we can rule out a value of 9 or more
(which all the above OB associations have) at the 99.8 percent
confidence level.  We stress that our ratio is worked out from the
spectroscopic follow-up of an optically-selected PMS sample, therefore
it is not biased against WTTS.  Furthermore we deliberately chose our
spectroscopic sample from a large area of the CMD to avoid any age bias.  

The explanation of such a low WTTS/CTTS value for Cep OB3b could
be ascribed to a less effective erosion (through stellar winds and
ionising radiation) of the circumstellar discs around the low-mass PMS
members by the high-mass members of the association.  
The kinematic age of $<1$ Myr found by a number of authors (see Section 1) 
probably indicates when much molecular material was lost and the association
became unbound, i.e., the bulk of the molecular material shielding the newly-born
low-mass stars from the dispersive influences of the high-mass stars
has only been recently cleared.
An interesting speculation is that this transition was
triggered by the supernova event centred on Cep OB3a that occurred less
than a million years ago (Simonson \& van Someren Greve 1976) and that
it terminated low-mass star formation.

The age of the low-mass PMS stars we have found in Cep OB3b is suggesting
disc lifetimes of about 5 Myr. 
Low-mass stars remain in the protostellar phase until an external
trigger obliges them to cross the birthline through the collapse of the
protostellar core and the dispersion of the protostellar envelope. At
that time they become optically visible and then, and only then, their
isochronal clock is set to 0.  At an age of about 1 Myr they are
visible as T Tauri stars: approximately half of them are still
surrounded by accretion discs (Strom, Strom \& Merrill 1993; Edwards
1993, and references therein) which can survive as long as 10 Myr (see
Strom et al. 1989; Skrutskie et al. 1990; Preibisch \& Zinnecker 1999).
Recently, Haisch, Lada \& Lada (2001a) reported the results of the
L-band survey of the clusters NGC 2264, NGC 2362 and NGC 1960 (of
intermediate age, 2.5-30 Myr), using JHKL colours to obtain a census of
the cluster objects with a circumstellar disc.  Coupled with previous
studies on other clusters, i.e, NGC 2024 (Haisch et al. 2001),
Trapezium (Lada et al. 2000) and IC 348 (Haisch, Lada \& Lada 2001b),
the results suggest that the disc fraction in a cluster is initially
very high ($> 80$ per cent), and rapidly decreases as the cluster age
increases; that about half the stars in the cluster lose their discs in
less than about 3 Myr; and finally that all the cluster members lose
their discs in about 6 Myr or more. 
All these papers use NIR excess as marker of circumstellar discs, but 
not all stars showing such an excess are strong H$\alpha$ emitters. 
The fact that we have detected the presence of accreting discs in 
50 per cent of TTS systems is a lower limit to the NIR excess disc fraction 
in Cep OB3b. The CTTS we have found are all less than 5 Myr old, 
consistent with the disc lifetime limits found by Haisch, Lada \& Lada (2001a).

\section{Conclusions}

Thanks to an optical CCD photometric survey, we have
discovered low-mass PMS stars in the younger subgroup of the Cep OB3
association, Cep OB3b.  Their presence has been confirmed by spectroscopic
follow-up of an optically-selected PMS sample. Out of 110 stars for
which we have spectra, we classified 10 PMS and 6 possible PMS objects,
all of which are kinematic members of the association.  Just 4 of the
PMS objects have {\it ROSAT} X-ray counterparts, suggesting the
presence of a substantial, less X-ray-active population.
Thus the {\it ROSAT} X-ray selected sample is likely to be highly
incomplete. Forthcoming, more sensitive {\it Chandra} observations will
probably uncover a much larger X-ray emitting population.

Among the PMS members, we found 5 CTTS and 5 WTTS with likely ages ranging
from about $<1$ and nearly 10\,Myr and masses in the range $\sim 0.9 - 3.0$
M$_{\odot}$.  It is difficult to explain the apparent age spread of PMS
objects around the isochrones solely by means of unresolved binaries,
photometric errors, spreads in distance and extinction.  If
the objects which appear younger than 1 Myr are real Cep OB3b members,
the spread would be broad enough to confirm a real spread in ages
across the subgroup: a spectroscopic follow-up for the very young
candidate association members is therefore needed to investigate this
and confirm that low-mass star formation has only recently ended in Cep OB3b.

The fact that there seem to be objects with photometric ages younger than 1 Myr 
in the westernmost fields of the subgroup could be explained as the result of 
star formation triggered by a supernova explosion. Indeed there is evidence 
for a SN remnant in the older subgroup of the association that happened closer 
to the easternmost fields. This would explain why star formation in the 
westernmost fields appears to be ended later.  
Another possible explanation would be the proximity of the easternmost fields 
to the major ionising source in the region -- the O7n star.

From the number of the newly discovered CTTS and WTTS the WTTS/CTTS
ratio is equal to 1.  Despite the presence of OB type members, this
value is closer to that found in T associations (1-13) than to that
found in OB associations (up to 20).  This suggests that the stellar
winds and ionising radiation from the high-mass members of the
association were less effective in eroding and subsequently evaporating
the circumstellar discs around the low-mass members of Cep 0B3b. We
speculate that it is probably due to the quite recent dispersal of
shielding molecular material in the association.

Further spectroscopy is urgently required to test these scenarios. In
addition, the forthcoming availability of 2MASS near infrared data over
the whole of our surveyed field will enable us to better isolate the
PMS populations and put further constraints on the presence and
timescales for dissipation of circumstellar material (Pozzo et al. in
preparation).

\section*{Acknowledgements}

This research has made use of the US Naval Observatory A2.0 catalogue.
The Isaac Newton Telescope is operated on the island of La Palma by the
Isaac Newton Group in the Spanish Observatorio del Roque de los
Muchachos of the Instituto de Astrofisica de Canarias.  MP was
supported by a Departmental Research Studentship in Keele University
during her PhD and is currently supported by a PPARC Postdoctoral
Research Fellowship at Imperial College. TN was in receipt of a PPARC
advanced fellowship when the majority of this work was carried out. We
thank the referee, Leisa Townsley for a very thorough report.

\appendix

\section{Cross-correlation with Jordi's catalogue} 

The majority of the works in the literature on the Cep OB3 association, are studies 
of the BHJ stars classified by BHJ. Besides these, there are some authors who studied 
the Cep OB3 region from a general point of view. 
S\"arg \& Wramdemark (1970) presented photoelectric photometry of early-type stars 
in a Milky Way field in Cepheus, and Jordi et al. (1992) performed Str\"omgren 
photometry for 45 stars in the association. We have not tried a cross-correlation 
with their catalogues, since all their observed stars are brighter than ours 
(respectively $V< 11.54$ and $V< 10.72$, whereas we have stars with V$>12$).   

We cross-correlated our optical catalogue with the catalogue of Jordi et al. 1995 
(hereafter J95). They observed 1056 stars from 18 randomly selected fields in Cep OB3, 
7 in the old and 11 in the young subgroup. Their catalogue contains
stars with $V= 8-21$ mag, with complete photometry in all the colours down
to $V=15.5$ for just 130 stars.

\begin{figure}
\vspace{6.3cm}
\includegraphics{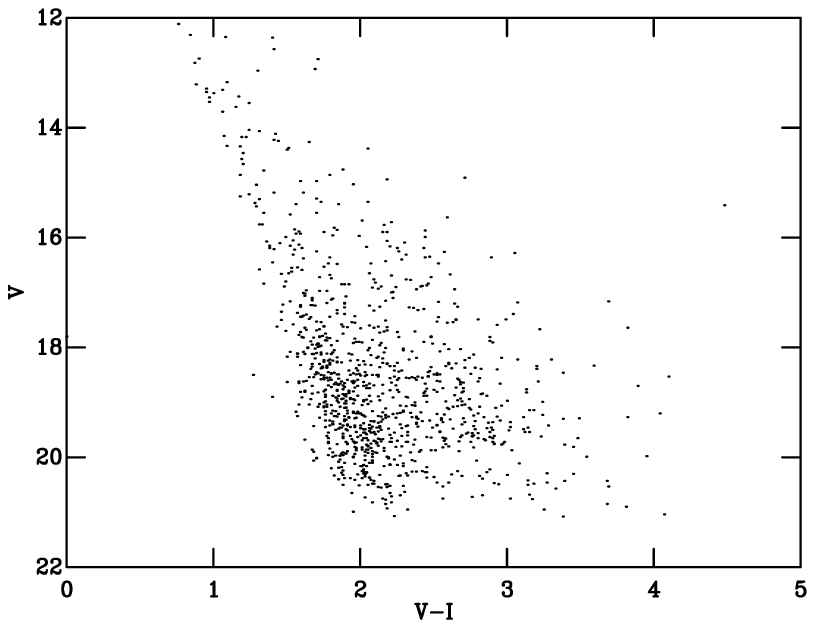}
\caption[The V versus (V-I) colour-magnitude plot for data obtained by J95]
{The V versus (V-I) colour-magnitude plot for data obtained by J95.}
\label{plotJ}
\end{figure}

In Fig. \ref{plotJ}, the V versus (V-I) colour-magnitude diagram derived from J95 is shown on 
the same scale as the one used in the plots of Figs. \ref{plot_9panels}. 
By comparing it with the CMD we have for just the field 1a (see Fig. \ref{plot_9panels}, 
top-left), we can see that the pre-main sequence is less defined, almost imperceptible. 
This is a dilution effect, due to the fact that their total number of stars is less than half 
the number we have in just one field, and that the objects belong to fields scattered all over 
the Cep OB3 region, with a field of view of just $3.0' \times 4.4'$ each.
Therefore, for the first time, thanks to a wide surveyed area and deep photometry, we have been
able to unambiguously detect a low-mass PMS population in the Cep OB3 association.

We cross-correlated our optical catalogue (just stars with zero-flags) with the J95 catalogue 
searching for correlations within an increasing radius. 
From the cumulative distribution of the number of matches 
obtained at different radii, the correlation radius at which we have obtained the maximum 
number of optical counterparts, before spurious correlations contributed significantly to the 
distribution, was found to be 1.85 arcsec (corresponding to 5 pixels on the CCD frames). 
Such big differences in positions are probably due to J95 astrometric errors, because our mean 
astrometric accuracy is about 0.4 arcsec. J95 used reference stars from the Guide Star 
Catalogue (GSC; with astrometric precision up to 0.8 arcsec; see Lasker et al. 1990), 
but they obtained an astrometric accuracy worse than 1 arcsec since the number of GSC 
reference stars falling on their fields is quite small. 
The limiting magnitude of the GSC in the 
Cep OB3 region is about V$=14$, therefore most of their fields contain just two 
reference stars.

\begin{figure}
\vspace{6.3cm}
\includegraphics{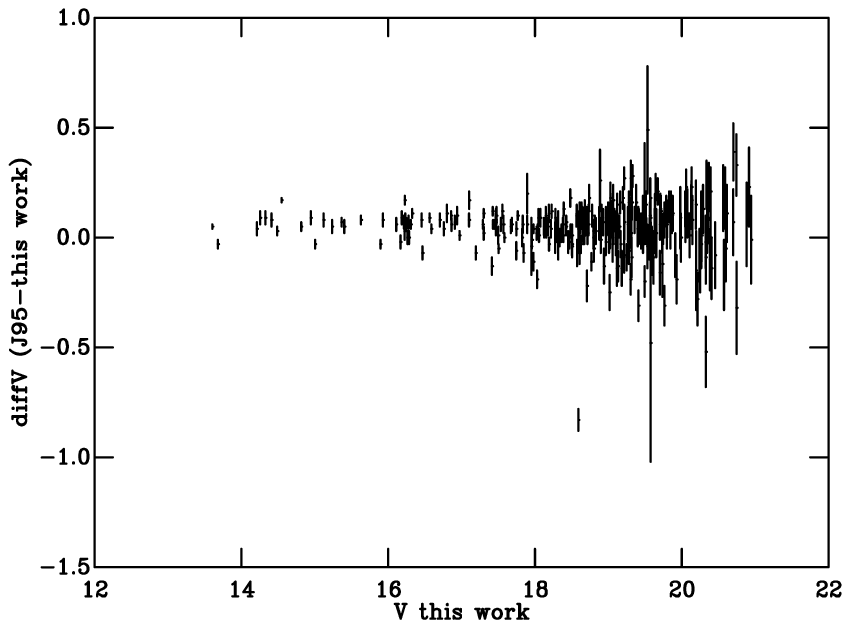}
\caption[V magnitude differences J95 - this work vs visual magnitudes from this work]
{V magnitude differences J95 - this work versus visual magnitudes
from this work for 302 matched stars. Error bars are the sum (in quadrature) of the errors 
for a given star from each catalogue.}
\label{VmVj}
\end{figure}

We found more than 300 matches in total (within the correlation radius of 1.85 arcsec), 
for stars in fields 1a, 2a, 2b, and 4b.
In Fig. \ref{VmVj} we compare our CCD measurements for V with the optical data given by J95, 
without restriction on the photometric errors: we find no systematic trend and a mean 
difference of about 0.04 mag. Analogously for (V-I), (B-V) and (U-B) colours we find 
mean differences of about 0.08, -0.04 and 0.06 respectively. 
The higher dispersion noticeable for fainter stars is an effect due in part 
to larger errors in the J95 catalogue, in part to some faint stars in our 
catalogue which are affected by light from a close saturated star, and thus have quite large
photometric errors. 
However, note that in this paper we performed our analysis using an optical catalogue
containing just good (i.e., with zero-flags) stars with photometric errors 
less than 0.1 mag. As an example, typical photometric errors in V, (B-V), (V-I) 
and (U-B) for stars in our catalogue at a given visual magnitude of V=19 for BVI 
colours and V=17 for (U-B) are respectively of $0.02, 0.05, 0.02$ and $0.02$.

\section{Additional extinction for the background population} 

In Section \ref{secstars} we have identified the objects plotted in the CMDs as belonging 
to a background sequence (the bluer) and a PMS (the redder), plus possible contamination 
from faint foreground stars.
We noticed (see Fig. \ref{plot_9panels}) that the background population in field 1a 
is also apparent as the blue sequence in each of fields 2a and 2b, whereas it is essentially 
absent from field 1b, positioned directly on the molecular cloud. 
In fields 3a, 3b, 4a and 4b (see Fig. \ref{plot_9panels}), the background population 
of stars is still apparent but there is not the neat separation from the PMS population.
Analogously, in Section \ref{isofit}, after isochrone fitting to the PMS objects, we noticed that 
in fields 1a and 2a there seems to be a neat separation between the two sequences (background 
and PMS): the 10-30 Myr gap is less populated. Instead in fields 2b and 3b we noticed no 
difference in the stellar densities moving from the 1 to 30 Myr isochrones.
This can be explained by additional interstellar material in these directions, the background 
objects being shifted towards larger (V-I) colours, contaminating the 10-30 Myr gap and part 
of the PMS.

\begin{figure}
\vspace{7cm}
\includegraphics{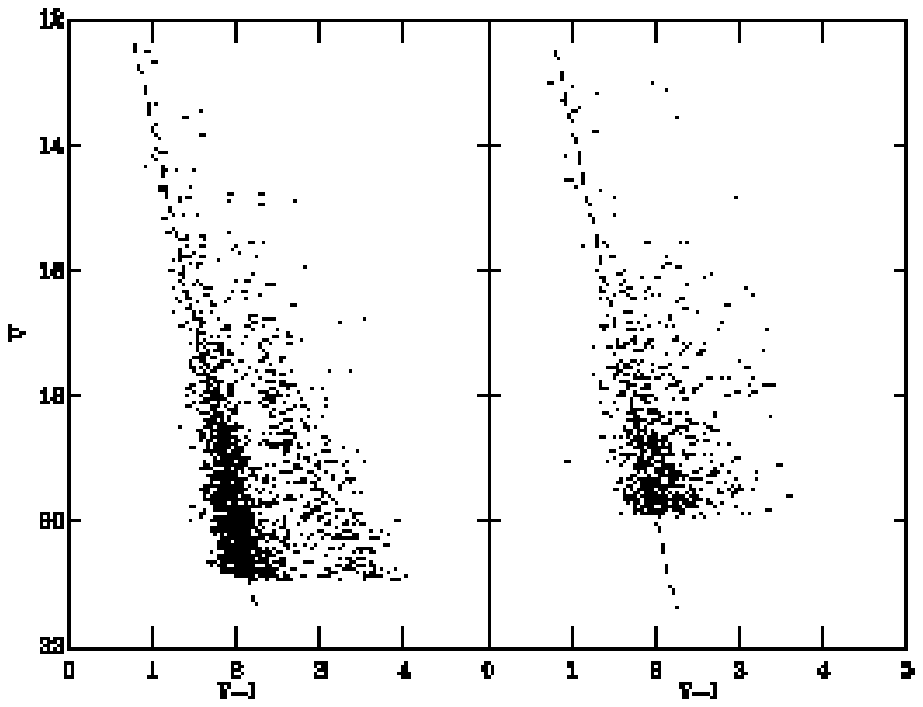}
\caption[Plot 1a - dered3a]
{CMD of field 3a [right], dereddened by the additional $A_V=0.6$ value (see text), 
necessary to match its background sequence with that of field 1a [left].}
\label{plot1a3a}
\end{figure}

In fact, we found that it is possible to match the background populations of fields other than 1a 
(i.e., 2a, 2b, 3a, 3b, 4a and 4b; note that for field 1b no comparison is possible because it is 
looking at the cloud and most of the background sequence is missing) with that of field 1a, 
shifting them up and to the left (i.e., dereddening them), along the reddening vector.
As an example, we have found that dereddening field 3a by an $A_V=0.6 \pm 0.2$ 
brings the background sequence into agreement with that in field 1a (see Fig. 
\ref{plot1a3a}). 
Analogously, it could be shown that the same argument holds for the other fields:
in their CMDs, the background sequences match that in field 1a once they have been dereddened 
by an additional extinction value which is about $A_V=0.15 \pm 0.10$, $A_V=0.6 \pm 0.2$, 
$A_V=0.5 \pm 0.2$, $A_V=0.5 \pm 0.1$, $A_V=0.8 \pm 0.2$ and $A_V=1.4 \pm 0.2$ 
for fields 2a, 3a, 4a, 2b, 3b and 4b respectively.

Therefore, the majority of the stars filling the gap between the 10 and 30 Myr isochrones 
in all the fields other than 1a are background stars which are shifted towards redder colours.
We think that this additional extinction is affecting just the background population
and that it is due to different concentrations of interstellar material in the direction of 
these fields: as confirmed in Fig. \ref{perseus}, there is a considerable assemblage of stars
beyond Cep OB3b and changing in the extinction values are perfectly plausible. 

Whereas this is certainly true for fields 2a, 2b and especially 3a and 4a, which are farther 
away from the Cep OB3 molecular cloud (see Fig. \ref{figfields3}), some doubts could be risen 
for 3b and 4b which are closer to the cloud: for these two fields, the molecular material could 
be responsible for most of the additional extinction we have determined. This also implies
that we cannot role out the presence of some heavily reddened (because they are more deeply 
embedded in the molecular material) PMS objects, but this would not alter the global appearance 
of the PMS locus in the CMDs of fields 3b and 4b.

\bsp

\label{lastpage}

\end{document}